\documentclass[aps,prfluids,twocolumn,showpacs,10pt,nobalancelastpage]{revtex4-2}
\usepackage[latin1]{inputenc}
\usepackage{graphicx}
\usepackage{amsmath}
\usepackage[dvipsnames]{xcolor}
\usepackage{times}
\usepackage{ulem}
\usepackage{enumitem}
\usepackage{bm}
\usepackage{hypernat}
\usepackage{hyperref}
\hypersetup
{
	colorlinks,%
	citecolor=blue,%
	linkcolor=blue,%
	urlcolor=blue,%
}

\begin{document}

\title{Effect of external tension on the wetting of an elastic sheet}

\author{Gregory Kozyreff}
\email{gregory.kozyreff@ulb.be}
\affiliation{Physics Department, Universit\unexpanded{\'e} libre de Bruxelles (ULB), CP231, 1050 Brussels, Belgium}
\author{Benny Davidovitch}
\email{bdavidov@umass.edu}
\affiliation{Department of Physics, University of Massachusetts, 01003 Amherst, Massachusetts, USA}
\author{S. Ganga Prasath}
\affiliation{School of Engineering and Applied Sciences, Harvard University, 02138 Cambridge, Massachusetts, USA}
\author{Guillaume Palumbo}\thanks{Present address: Laboratoire Interfaces \& Fluides Complexes, Universit\unexpanded{\'e} de Mons, 20 Place du Parc, Mons, 7000, Belgium}
\affiliation{Nonlinear Physical Chemistry Unit, Universit\unexpanded{\'e} libre de Bruxelles (ULB), CP231, 1050 Brussels, Belgium}
\author{Fabian Brau}
\email{fabian.brau@ulb.be}
\affiliation{Nonlinear Physical Chemistry Unit, Universit\unexpanded{\'e} libre de Bruxelles (ULB), CP231, 1050 Brussels, Belgium}

\begin{abstract}
Recent studies of elasto-capillary phenomena have triggered interest in a basic variant of the classical Young-Laplace-Dupr\'e (YLD) problem: The capillary interaction between a liquid drop and a thin solid sheet of low bending stiffness. Here, we consider a two-dimensional model where the sheet is subjected to an external tensile load and the drop is characterized by a well-defined Young's contact angle $\theta_Y$. Using a combination of numerical, variational, and asymptotic techniques, we discuss wetting as a function of the applied tension. We find that, for wettable surfaces with $0<\theta_Y<\pi/2$, complete wetting is possible below a critical applied tension thanks to the deformation of the sheet in contrast with rigid substrates requiring $\theta_Y=0$. Conversely, for very large applied tensions, the sheet becomes flat and the classical YLD situation of partial wetting is recovered. At intermediate tensions, a vesicle forms in the sheet, which encloses most of the fluid and we provide an accurate asymptotic description of this  wetting state in the limit of small bending stiffness. We show that bending stiffness, however small, affects the entire shape of the vesicle. Rich bifurcation diagrams involving partial wetting and ``vesicle'' solution are found. For moderately small bending stiffnesses, partial wetting can coexist both with the vesicle solution and complete wetting. Finally, we identify a tension-dependent bendo-capillary length, $\lambda_\text{BC}$, and find that the shape of the drop is determined by the ratio $A/\lambda_\text{BC}^2$, where $A$ is the area of the drop.
\end{abstract}


\date{\today}
\maketitle

\section{Introduction}\label{sec:intro}

Elasto-capillary phenomena, namely mechanical deformations of elastic bodies due to capillary forces, are at the focus of a growing attention. Indeed, aside from fundamental interests, this field of research appears to be relevant to the study of budding in biological cells and other biomimetic systems~\cite{Long2007,Li2011} and opens new perspectives for fabrication at small scales where surface tension dominates volume weight. For example, capillary forces can be used to fold elastic sheets into desired three-dimensional objects~\cite{Syms2003,Py07,Guo09,Roman10,Pineirua10,Neukirch13,Brubaker16a,Brubaker16}. In this context, focus has been made on sheets with free ends. However, very thin sheets are often subjected to external tensile loads due to capillary forces for floating sheets~\cite{Huang07,Schroll13,Toga13} or due to clamped ends~\cite{Nadermann13,Hui15,Schulman15}. We thus propose to fill this gap by studying the influence of an applied external tension on the wetting states of a drop deposited on a thin elastic sheet.

\begin{figure}[!t]
\centering
\includegraphics[width=\columnwidth]{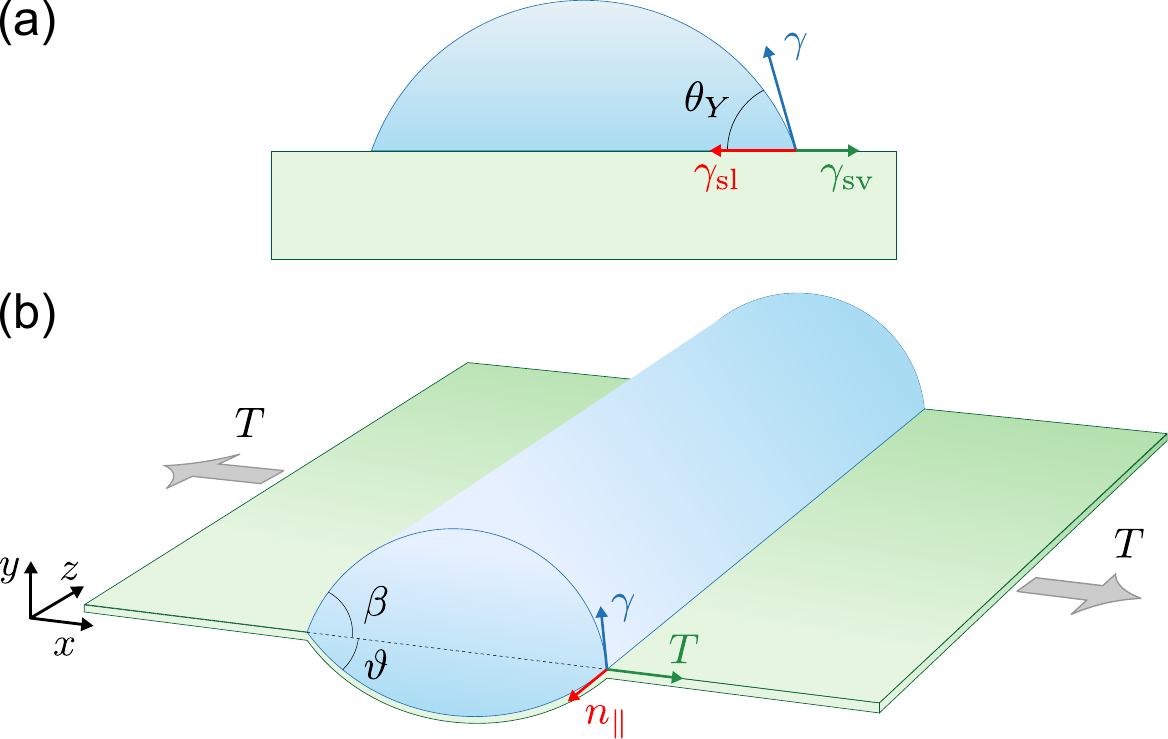}
\caption{(a) Classical Young-Laplace-Dupr\'e picture of partial wetting on a perfectly rigid substrate, involving liquid-vapor surface tension, $\gamma$, and forces associated with ``surface stress'', $\gamma_{\text{sv}}$, $\gamma_{\text{sl}}$, that act parallel to the solid surface. (b) Schematic of our model system. Additional control parameters: applied tension $T$ and bending modulus $B$.} 
\label{fig01-YLD}
\end{figure}

One of the basic questions in the study of elasto-capillary phenomena is how the wetting states are modified, if the underlying assumption of a perfectly rigid, semi-infinite solid substrate is relaxed, see Fig.~\ref{fig01-YLD}. When the solid substrate is undeformable, the wetting states of a given volume of liquid are determined by the Young-Laplace-Dupr\'e (YLD) equation~[Fig.~\ref{fig01-YLD}(a)]
\begin{align}
\cos \theta_Y &= \Delta\gamma/\gamma, &\Delta\gamma&\equiv\gamma_{\text{sv}} -\gamma_{\text{sl}},
\label{eq:YLDlaw}
\end{align}
where $\theta_Y$ is the contact angle between the solid-liquid and liquid-vapour interfaces whereas $\gamma_{\text{sv}}$, $\gamma_{\text{sl}}$, and $\gamma$, are respectively, the solid-vapor, solid-liquid, and liquid-vapour surface energies (see for example \cite{Neukirch13}). When $\Delta\gamma<-\gamma$, surface energy disfavours any liquid-solid contact (non-wetting). Conversely, when $\Delta\gamma>\gamma$, the system is said to be in a complete wetting state. For intermediate values, $-\gamma<\Delta\gamma<\gamma$, that is $0<\theta_Y<\pi$, the system is in a partial wetting state.

When the solid substrate is deformable (low elastic Young's modulus $E$) and thick (unbendable), the local deformation of the solid surface is on the order of the elastocapillary length $\ell_\text{EC} = \gamma/E$, and the relevant dimensionless parameters are the ratios, $\ell_\text{EC}/a$, and $\ell_\text{EC}/R$, where $a$ is a microscopic (atomic or molecular) length, and $R$ is the characteristic drop size. The rich physics that emerges at various ranges of theses dimensionless parameters has been the subject of theoretical works~\cite{Rusanov75,Shanahan87b,Marchand12,Style12,Hui14,Lubbers14,Dervaux15,Cao15,Andreotti16,Dervaux20,Pandey20} and experiments~\cite{Yuk86,Extrand96,Pericet08,Jerison11,Style13,Kim2021} (see also some recent reviews~\cite{Style17,Andreotti20}).

When the solid substrate is a thin, bendable sheet, another length scale comes into consideration: the bendo-capillary length $\ell_\text{BC}=\sqrt{B/\gamma}$, where $B=Et^3/12(1-\nu^2)$ is the bending stiffness~\cite{Howell09}. The plate is then easily bent by a liquid drop when $\ell_\text{BC}$ is small compared to the size of the drop, i.e. when the \textit{bendability parameter}
\begin{equation}
\label{eq:cap-bend}
\Gamma\equiv R^2\,\ell_\text{BC}^{-2}=\gamma\, R^2\, B^{-1}
\end{equation}
is large. In addition, the local surface deformation is small when $\ell_\text{EC}/t\ll 1$. This is the situation that we consider in the present study. In terms of $t$, we are thus focusing on the double limit $\ell_\text{EC} \ll t \ll R^{2/3} \ell_\text{EC}^{1/3}$. The far edges of the sheet, away from the liquid drop, may be free \cite{Py07,Guo09,Roman10,Pineirua10,Neukirch13,Brubaker16}, clamped~\cite{Nadermann13,Hui15,Schulman15}, or subject to a fixed tensile load by a liquid sub-phase~\cite{Huang07,Schroll13,Toga13}. 

The parameter regime $\Gamma \gg 1$ and $\ell_\text{EC}/t \ll 1$ describes sheets that are ``highly bendable'' yet ``nearly inextensible''. In this regime, there is a large contrast between the bending energy, $U_{\text{bend}}$, and the strain energy, $U_{\text{strain}}$, whose high cost must be taken into consideration either explicitly~(when studying finite liquid volume, in which case the liquid drop imposed Gaussian curvature on the sheet~\cite{Shanahan87,Olives93,Schroll13,Davidovitch18}), or by imposing an inextensibility constraint, as we will do in this paper for a simplified model system.

From a practical purpose, this parameter regime characterizes a large range of solid sheets that are commonly studied in the material science community: from common elastomers ($E \sim$ MPa) with a thickness of few micrometers to stiff polymers ($E \sim$ GPa) with a thickness of few hundreds of nanometers. For a characteristic drop size ranging from few tens to few hundreds of micrometers~\cite{Schulman15}, $\Gamma$ varies roughly between $1$ and $10^6$ whereas $\ell_\text{EC}/t$ can be as small as $10^{-5}$. Most experiments reported in Refs.~\cite{Nadermann13,Schulman15,Schulman17,Adam17,Huang07,Schroll13,Toga13} are in this parameter regime.

In this paper, we consider a two-dimensional model composed of a liquid cylinder of cross-sectional area $A \equiv R^2$ in contact with a rectangular solid sheet of length $L \gg R$ under an applied tension $T$, see Fig.~\ref{fig01-YLD}(b). We further assume that gravity is negligible. The absence of Gaussian curvature considerably simplifies the analysis compared to more realistic 3D problems~\cite{Huang07,Kusumaatmaja2011,Schroll13,Brubaker16} and allows us to push the analytical investigation beyond scaling laws. Notice however that such a two-dimensional system can, to some extent, be realized experimentally using a thin elastic filament floating on a fluid surface and wet by a droplet of another immiscible fluid~\cite{Prasath2021}. In contrast to previous works on a related system~\cite{Py07,Neukirch13,Brubaker16a,Brubaker16}, we pay close attention to the effect of tensile loads, $T$, exerted on the solid sheet at its far edges, away from the liquid drop and thus, we consider the effect of another dimensionless parameter, $T/\gamma$, in addition to $\Gamma$. Numerically, we study the system for both small and large $\Gamma$. Analytically, we treat the high-bendability limit, $\Gamma \gg 1$ by singular perturbation theory. Most of our results are derived in the wettable regime, i.e. $0<\theta_Y<\pi/2$. 

\begin{figure}[!t]
\centering
\includegraphics[width=\columnwidth]{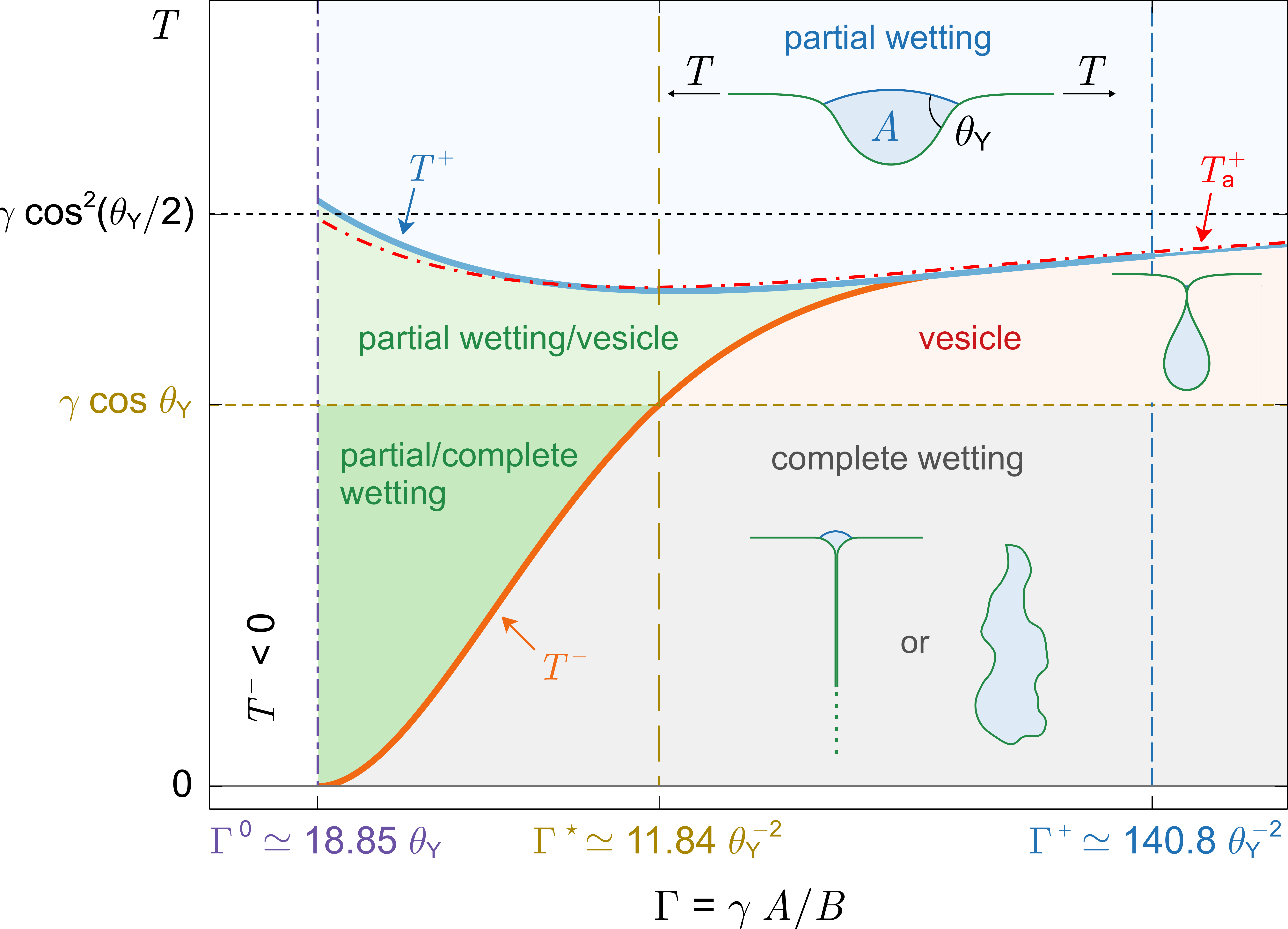}
\caption{A ``wettability phase diagram'' of our 2D model, for a given value of $0<\theta_Y<\pi/2$, exhibits three types of energetically-favourable states of a liquid drop of area $A$ in contact with an elastic sheet of length $L\gg\sqrt{A}$ and bending modulus $B$. The control parameters are the applied tension, $T$, and the capillary bendability parameter $\Gamma$, Eq.~(\ref{eq:cap-bend}). The dot-dashed red line $T^{+}_a$ is a curve derived from the asymptotic analysis (Sec.~\ref{sec:partialwrap}) of the vesicle that approximates the numerical boundary $T^{+}$ of existence of the vesicle state. The vesicle and partial wetting states are illustrated by numerical solutions; the complete wetting state is sketched for either finite or infinite $L$.}
\label{fig-phase-diagram}
\end{figure}

Our results are succinctly summarized in the schematic phase diagram, Fig.~\ref{fig-phase-diagram}, on which we briefly elaborate below:

\begin{enumerate}
\setlength\itemsep{0em}
\item As $T\to\infty$ (i.e., $T\gg \gamma$), the sheet becomes asymptotically flat and is only partially wet, with a contact angle given by the classical value, Eq.~(\ref{eq:YLDlaw}). Such a \textit{partial wetting} state persists for all tensions $T>T^{-}\left(\theta_Y,\Gamma\right)$ and is the unique wetting state when $T>T^{+}\left(\theta_Y,\Gamma\right)$.
\item In the range $\cos\theta_Y<T<T^{+}\left(\theta_Y,\Gamma\right)$ the sheet can be in self-contact and form a \textit{vesicle} that wraps most (but not all) of the liquid. The two curves $T^{\pm}\left(\theta_Y,\Gamma\right)$ merge at the point $\Gamma=\Gamma^+(\theta_Y)$ and in the limit $\Gamma\to\infty$, they asymptote to $\gamma \cos^2\left(\theta_Y/2\right)$. Analytically, we show that the shape of the vesicle is uniquely determined by the value of the product $\Gamma(T -\gamma \cos \theta_Y)$, a result that is found to hold even for $\Gamma=O(1)$. We derive the approximate curve $T_a^{+}\left(\theta_Y,\Gamma\right)$, in very good agreement with the numerics.
\item Finally, $T < \gamma\cos\theta_Y$ is the range of existence of the \textit{complete wetting state} for a sufficiently bendable sheet ($\Gamma  > \Gamma^{\star}(\theta_Y)$), whereby the liquid completely wets one side of the sheet. Recalling that in the classical YLD picture complete wetting is obtained only if $\theta_Y= 0$, we see that high bendability enables a complete wetting state even if $\theta_Y > 0$, provided the tensile load is sufficiently small.
\item While the vesicle and complete wetting state are mutually exclusive, the light and dark green regions in Fig.~\ref{fig-phase-diagram} indicate that the partial wetting state can coexist with the former if $\Gamma<\Gamma^{+}(\theta_Y)$ and with both if $\Gamma<\Gamma^{\star}(\theta_Y)$, paving the way to hysteresic behaviour.
\end{enumerate}

Thus, for a given value of $0<\theta_Y <\pi/2$, the partial wetting predicted by the classical YLD law for non-bendable solids separates into three distinct phases -- complete wetting, vesicle, and partial wetting -- enabled by the floppiness of the solid.

The paper is organized as follow. In Sec.~\ref{sec:zero-B}, we set the stage by discussing the limit of zero bending stiffness ($\Gamma = \infty$). For a given $\theta_Y$, we identify the three asymptotic wetting states at distinct intervals of $T/\gamma$: complete and partial wetting and an intermediate wetting state where the sheet forms a circular vesicle and wraps the entire liquid area (see Fig.~\ref{fig03-bifurcation-nb}).


The inclusion of bending stiffness starts in Sec.~\ref{sec:finite-B}, where the governing equations of the system are presented. These are studied numerically in Sec.~\ref{sec:num} where we show how the three wetting states identified in Sec.~\ref{sec:zero-B} occupy distinct regions in the parameter space spanned by $T$, $\Gamma$ and $\theta_Y$. 
Furthermore, we show that the shape of the vesicle can be completely altered for any $\Gamma < \infty$. 
Section~\ref{sec:partialwrap} is devoted to the asymptotic analysis of the ``vesicle'' state in the limit $\Gamma \gg 1$. Finally, we conclude in Sec.~\ref{sec:conclusion}. 

\section{Inextensible, infinitely bendable sheet}\label{sec:zero-B}

In the 2D model considered here, the energy of the sheet is given by: 
\begin{equation}
U  = U_s + \mathcal{W}  + U_{\text{elas}},
\end{equation}
and comprises a surface energy $U_s$, a work $\mathcal{W}$ (done by pulling the edges) and an elastic energy $U_{\text{elas}}$ due to the deformation of the sheet with respect to its strainless, planar shape. The energetically costly stretch is eliminated by the inextensibility constraint, whereas the bending cost is expected to be small in comparison to the surface energy for thin enough sheets. Hence, we start in this section by ignoring the bending stiffness altogether. 

\begin{figure}[!t]
\centering
\includegraphics[width=\columnwidth]{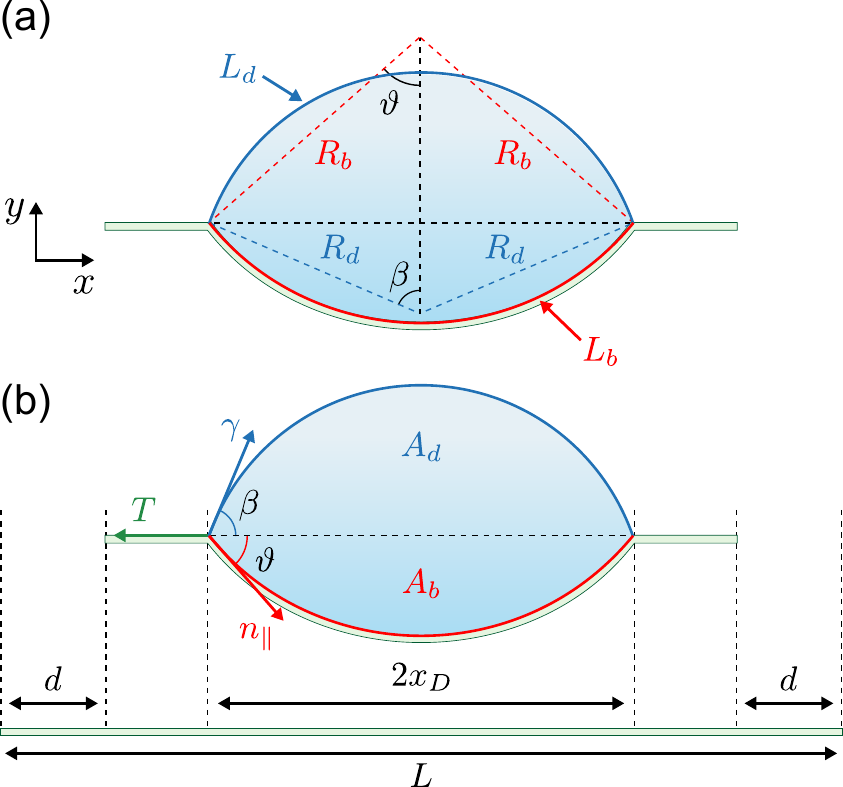}
\caption{Schematics of a cross section of the system for an infinitely bendable sheet ($B=0$). (a) The radii of curvature, $R_b$, of the wet part of the sheet and, $R_d$, of the liquid-vapour interface, and the relations with the lengths of the respective circular segments, $L_b=2\vartheta R_b$ and $L_d=2\beta R_d$. (b) The displacement, $2d=L_b-2x_D$, with respect to a flat sheet prior to wetting, where $x_D=R_b \sin \vartheta$ is the projection of the liquid-vapour interface onto the $x$-axis. $A_d$ and $A_b$ are the liquid areas enclosed between the chord of length $2x_D$ and the liquid-vapour and the wet part of the sheet, respectively.}
\label{fig02-displacement}
\end{figure}

\begin{figure*}[!t]
\centering
\includegraphics[width=\textwidth]{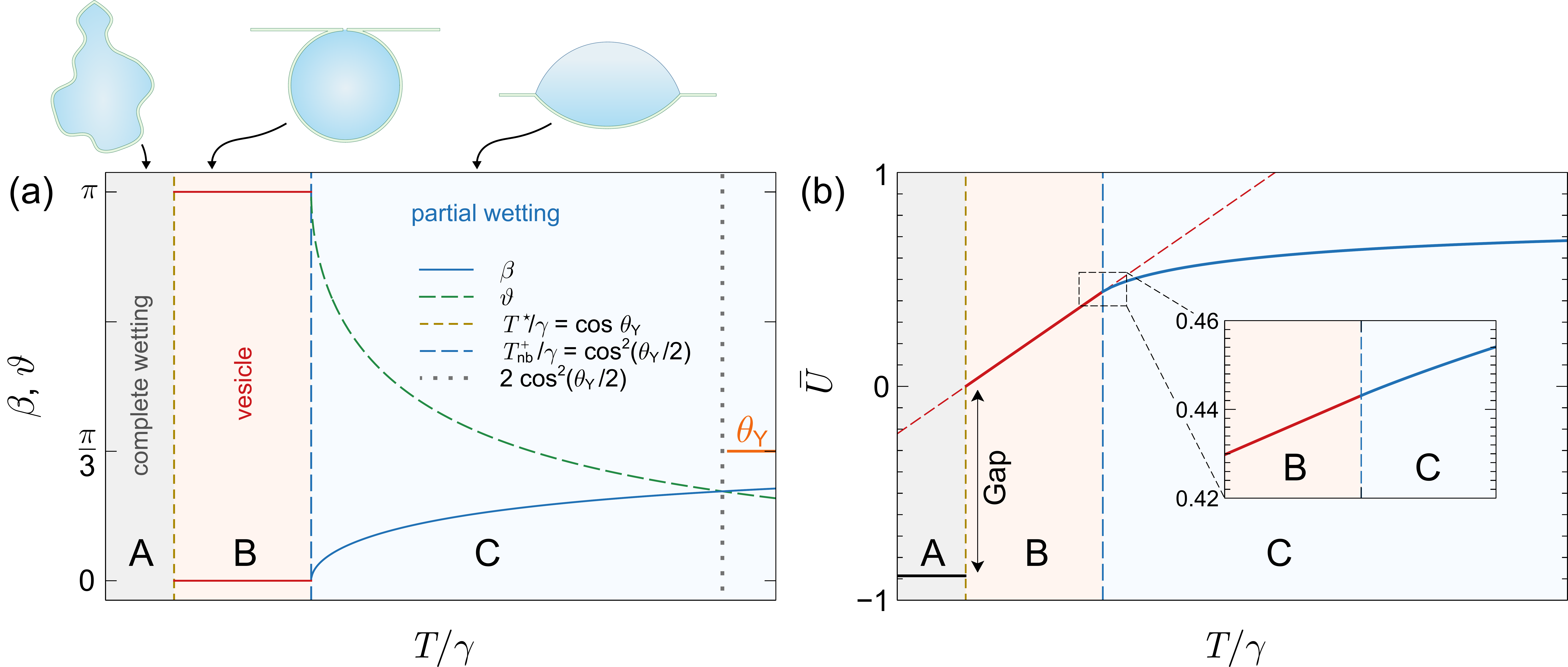}
\caption{The three wetting states in the infinite bendability limit (A: complete wetting; B: vesicle; C: partial wetting). 
(a) Dependence of the angles, $\beta$ and $\vartheta$, on the rescaled applied tension, $T/\gamma$, for $\theta_Y=\pi/3$. (b) Reduced energy $\bar U =(U-2\gamma_\text{sv}L)/(2\gamma\sqrt{A})$ as a function of $T/\gamma$ for $\theta_Y=\pi/3$. The black curve shows $\bar{U}_{\text{cw}}$, Eq.~(\ref{eq:Energywet}), for $L = 2\pi R_b$. The energy and its first derivative are both continuous at the partial wetting-vesicle transition. By contrast, the transition between complete wetting (A) and vesicle (B) is discontinuous, characterized by an energy gap at $T = T^{\star} = \gamma\, \cos \theta_Y$.} 
\label{fig03-bifurcation-nb}
\end{figure*}

As stated already in the introduction, we consider an infinitely long rectangular sheet in contact with a cylindrical drop, whose cross-sectional area is $A \equiv R^2$. Upon making contact with the drop, the wet part of the sheet becomes bulged, with a constant radius of curvature, $R_b$, due to the Laplace pressure in the drop, $p = \gamma/R_d$, where $R_d$ is the constant radius of curvature of the liquid-vapor interface, see Fig.~\ref{fig01-YLD}(b) and Fig.~\ref{fig02-displacement}.

In the absence of bending stiffness, the system energy is the sum of the surface energy and the work done by tensile loads at the edges of the sheet. The surface energy reads as
\begin{align}
\label{surf-ener}
U_s &= \gamma\, L_d + \gamma_{\text{sl}}\, L_b + (2L-L_b)\,\gamma_{\text{sv}} \nonumber \\ 
&= \gamma L_d -\Delta \gamma\, L_b + 2L \, \gamma_{\text{sv}},
\end{align}
where $L$ is the total length of the sheet, whereas $L_b$ and $L_d$ are, respectively, the length of the bulged part of the sheet and of the liquid-vapour interface, see Fig.~\ref{fig02-displacement}(a). The work performed by $T$ is
\begin{equation}
\mathcal{W} = 2d\, T=(L_b-2x_D)\, T,
\end{equation}
where $2x_D$ is the projected length of the liquid-vapour interface along the horizontal $x$-axis. Since the liquid-vapour and solid-liquid interfaces are necessarily circular, we have the following geometrical relations
\begin{subequations}
\label{geometry-zero-B}
\begin{align}
L_d = 2\beta\, R_d, \quad
L_b=2\vartheta\, R_b, \\
\label{geometry-zero-B-xd}
x_D=R_d\, \sin \beta=R_b\, \sin \vartheta,
\end{align}
\end{subequations}
where the last relation indicates that the two circular segments share the same chord, see Fig.~\ref{fig02-displacement}. The total energy $U_\text{pw}=U_s + \mathcal{W}$ of the partial wetting state is then given by
\begin{equation}
\label{energy}
U_\text{pw} = 2\gamma\, \beta\, R_d -\Delta \gamma\, L_b + 2L \, \gamma_{\text{sv}}+(L_b-2R_d\, \sin \beta)\, T,
\end{equation}
where the quantities $x_D$, $R_b$ and $L_d$ have been expressed in terms of $R_d$, $\beta$ and $\vartheta$. To minimize the total energy under the constraints of (i) fixed transverse area of the drop, $\mathcal{A}(\beta,\vartheta,R_d) = A$ [see Fig.~\ref{fig02-displacement}(b)], with $\mathcal{A}$ given by 
\begin{equation}
\mathcal{A} = R_d^2\left(\beta-\frac12\sin2\beta\right)+R_b^2\left(\vartheta-\frac12\sin2\vartheta\right),
\label{nobending:eq:A}
\end{equation}
where $R_b = R_d \, \sin \beta/\sin \vartheta$, and (ii) the geometric relation (\ref{geometry-zero-B}) for $L_b$, we introduce the functional
\begin{equation}
\label{lagrangian-no-B}
\mathcal{L}(\beta,\vartheta,R_d,L_b) = U_\text{pw} + \mu \left(A-\mathcal{A}\right) - \eta\left(L_b-\frac{2\vartheta R_d \sin \beta}{\sin \vartheta}\right),
\end{equation}
where $\mu$ and $\eta$ are Lagrange multipliers, that correspond, respectively, to the pressure $p$ in the liquid drop and the  parallel traction $n_{\parallel}$ in the wet part of the sheet (which for $B=0$ identifies with the tension in the sheet). The equilibrium equations are found by minimizing $\mathcal{L}$ with respect to $\beta$, $\vartheta$, $R_d$, and $L_b$, see Appendix~\ref{app-mini}. We find $\mu \equiv p = \gamma/R_d$ and $\eta \equiv n_{\parallel}$, with
\begin{subequations}
\label{eqs-equi}
\begin{align}
\label{eqs-eta1}
n_{\parallel} &= \gamma \frac{\sin \beta}{\sin \vartheta}, \quad T =\gamma\, \cos \beta + n_{\parallel} \, \cos \vartheta, \\
\label{eqs-eta2}
n_{\parallel} &= T-\Delta \gamma = T- \gamma\, \cos \theta_Y.
\end{align}
\end{subequations}
Equations (\ref{eqs-eta1}) are simply the vertical and horizontal force balance at the contact line, displayed in Fig.~\ref{fig02-displacement}(b). Equation (\ref{eqs-eta2}) is the familiar YLD law for the stress jump at the contact line. Solving these equations leads to
\begin{subequations}
\label{beta-theta-exact-inex}
\begin{align}
\label{beta-exact-inex}
\cos \beta &= \cos\theta_Y+ \frac{\gamma\sin^2\theta_Y}{2T}, \\
\label{theta-exact-inex}
\cos \vartheta &= 1- \frac{\gamma^2\sin^{2}\theta_Y}{2 T\left(T-\gamma\cos\theta_Y\right)},
\end{align}
\end{subequations}
We call the state given by Eqs.~(\ref{beta-theta-exact-inex}) the \textit{partial wetting state} because, in the limit $T/\gamma\to\infty$, it tends to the classical solution of a drop on a semi-infinite rigid substrate: $\vartheta \to 0$, $\beta\to\theta_Y$. Equations~(\ref{beta-theta-exact-inex}) implies that a partial wetting state exists only if
\begin{equation}
\label{Tnb-def}
T \ge T^{+}_{\text{nb}} =\gamma \cos^2\left(\theta_Y/2\right),
\end{equation} 
where the subscript `nb' stands for `no-bending limit'. Interestingly, the symmetric situation $\beta=\vartheta$ is obtained exactly at $T=2T^{+}_{\text{nb}}$.

The condition~(\ref{Tnb-def}) does not have an analog in the classical YLD theory of a drop on a thick (unbendable) solid body; it defines a minimal tensile load that is necessary to maintain a partial wetting contact even if $0<\theta_Y<\pi/2$. As $T \to T^{+}_{\text{nb}}$, $\beta\to0$ and $\vartheta \to \pi$ such that the wet part of the sheet tends to a closed circle and wraps the entirety of the fluid. Such a circular shape satisfies the conditions of static equilibrium at all tensions below that threshold. Henceforth, we call it the \textit{vesicle state}. Equations~(\ref{beta-theta-exact-inex}) implies also that the partial wetting state emerges supercritically from the vesicle state at $T^{+}_{\text{nb}}$ when the applied tension increases. However, Eqs.~(\ref{beta-theta-exact-inex}) do not determine whether the circular vesicle is stable for all $T< T^{+}_{\text{nb}}$.

One way to address this issue is to introduce a finite, arbitrarily small amount of bending stiffness as we do from Sec.~\ref{sec:finite-B} onwards. This rounds off the corner in the elastic sheet near the triple line [see inset of Fig.~\ref{fig04-setup}(a)] and makes Eq.~(\ref{eqs-eta2}) appears as the true force balance at the triple line. From this perspective, the force balance equations (\ref{eqs-eta1}) hold at a distance of a few elasto-capillary lengths away from the triple line where the angles $\vartheta$ and $\beta$ can be measured. Therefore, the contact angle $\vartheta+\beta$ is only the ``apparent'' contact angle, as measured in Ref.~\cite{Schulman15,Schroll13}, but the true contact angle as measured at a distance smaller than $\ell_{BC}$ from the contact line remains Young's angle, $\theta_Y$, in agreement with recent experiments performed on a related system~\cite{Twohig2018}. 

\subsection{The vesicle state}
\label{sec:2-wrapped}

As we noted above, when $T < T^{+}_{\text{nb}}$, the sheet wraps the entirety of the liquid drop ($\beta=0$, $\vartheta = \pi$) so that the state is characterized by $L_d = 0$, $L_b = 2\pi R_b$, and $A = \pi\, R_b^2$, see Eqs.~(\ref{geometry-zero-B}) and (\ref{nobending:eq:A}). The energy (\ref{energy}) becomes
\begin{equation}
U_{\text{ves}} =2\gamma_\text{sv}L+2\sqrt{A\pi}\left(T-\gamma\cos\theta_Y\right).
\label{eq:Uwrapped}
\end{equation}
Let us compare this energy with the one of the partial wetting state, in the vicinity of the threshold $T^{+}_{\text{nb}}$. Expanding Eqs.~(\ref{energy}) and (\ref{beta-theta-exact-inex}) above the threshold $T=T^{+}_{\text{nb}}$, we find
\begin{equation}
\label{ener-total-inex-exp}
U_\text{pw}-U_\text{ves} =-\frac{32}{3 \sin \theta_Y}\sqrt{A}\gamma \epsilon^{3/2}+\mathcal{O}(\epsilon^{5/2}),
\end{equation}
where $0<\epsilon = T-T^{+}_{\text{nb}} \ll 1$. Hence the partial wetting state has lower energy than the vesicle state when $T>T^{+}_{\text{nb}}$. The continuity of the angles $\beta$ and $\vartheta$, in the vicinity of the transition between the two states, reflects the continuity of both the energy, $U = U_{\text{ves}}$, and its first derivative, $dU/dT =dU_{\text{ves}}/dT$, at $T=T^{+}_{\text{nb}}$. Hence, in the infinite bendability limit, the transition is a continuous, second order transition. Specifically, we have a pitchfork bifurcation whereby the angles $\beta$ and $\vartheta$ vary rapidly with a small increase of the applied force past the bifurcation point. We will see in Sec.~\ref{sec:num} how adding bending energy to the model affects the nature of the transition.

\subsection{Complete wetting}
\label{sec:2-bendability}

We now revisit the assumption that the drop shape consists of circular segments. If there is a finite liquid-vapour interface, Laplace's law implies that it is necessarily a circular arc as well as the rest of the drop's interface which makes a contact with the sheet. However, for $T < T^{+}_{\text{nb}}$, the liquid in the vesicle state does not have a finite contact length with the vapour. Hence, we must address the possibility that the drop, once fully wrapped by the sheet, is no longer circular. 

For this purpose, we consider the energy of a vesicle of perimeter $L_b$ whose shape is not necessarily circular:
\begin{equation}
2\gamma_{\text{sv}}L + (T -\gamma \cos \theta_Y) L_b.
\end{equation}
For a circular vesicle, $L_b = 2\sqrt{A\pi}$ but otherwise $2\sqrt{A\pi} <L_b \le L$. Therefore, if 
\begin{equation}
\label{tnb-star-def}
T < T^{\star}=\gamma \cos \theta_Y,
\end{equation}
the energy is minimal for $L_b=L$, i.e. the liquid wets the entire length of the sheet. As a consequence, for $0<\theta_Y <\pi/2$, a tensile load $T < \gamma \cos\theta_Y$ is not sufficient to stabilize the sheet against a complete wetting by the drop. The energy of the complete wetting state is
\begin{equation}
U_\text{cw}=\left(\gamma_\text{sl}+\gamma_\text{sv}\right)L=2\gamma_\text{sv}L-\gamma\cos\theta_YL.
\label{eq:Energywet}
\end{equation}
A comparison between Eqs.~(\ref{eq:Energywet}) and (\ref{eq:Uwrapped}) shows that, at $T^{\star}$, the system undergoes a discontinuous transition, characterised by a finite energy gap, see Fig.~\ref{fig03-bifurcation-nb}(b). This gap can be viewed as a potential barrier to the formation of a vesicle from a complete wetting state which can only be overcome by applying a sufficiently large force on the sheet's edges ($T \ge \gamma \cos\theta_Y$). Notice that, when $\pi/2 < \theta_Y < \pi$ such a transition requires a compressive force ($T < \gamma \cos\theta_Y <0$) and complete wetting is therefore unobservable under any (or none) tensile load.

\section{Finite bendability: Model equations}\label{sec:finite-B}

\begin{figure}[!t]
\centering
\includegraphics[width=\columnwidth]{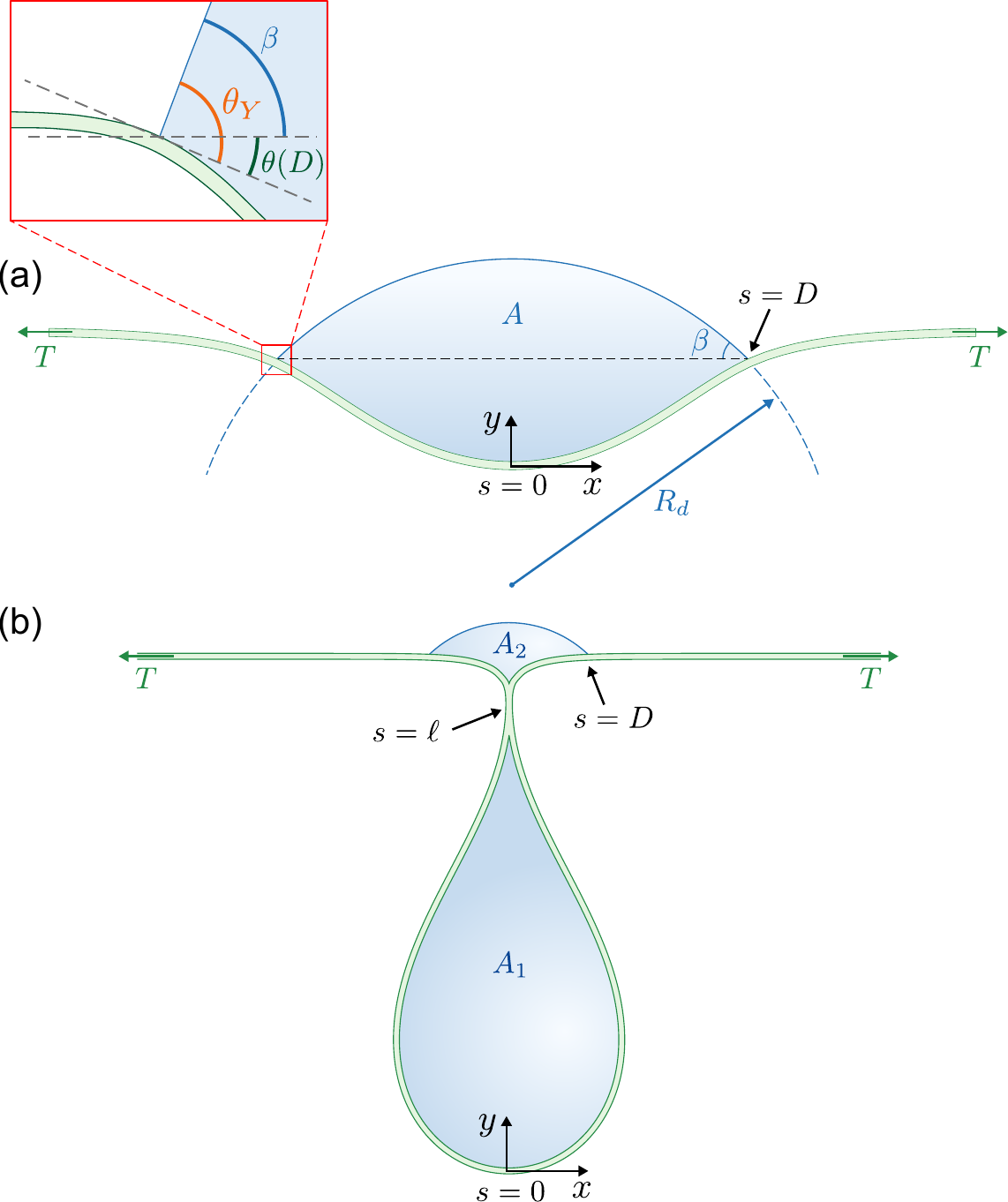}
\caption{Schematics of the system for (a) the partial wetting state and (b) the vesicle state.}
\label{fig04-setup}
\end{figure}

We now consider an elastic sheet with a bending modulus $B>0$ and set up the mathematical model to describe the partial wetting and vesicle state that are schematically depicted in Fig.~\ref{fig04-setup}. By symmetry, we may restrict our attention to $x \ge 0$.

\subsection{Partially wet state}

Denoting by $\kappa$, $n_\perp$, and $n_\parallel$ the curvature, perpendicular and parallel tractions along the elastic sheet, respectively, the local balance of forces and torques are, in the absence of self contact (see, e.g., \cite{Djondjorov2011,Marple2015} or \cite[p. 189]{Howell09}), 
\begin{align}
\label{eq:nperp1}
&B\, \partial_s\kappa =-n_\perp, &\partial_s n_\perp &=p-\kappa\, n_\parallel, &\partial_s n_\parallel &=\kappa\, n_\perp,
\end{align}
where $s$ is the distance along the sheet and $\partial_s$ denotes derivative with respect to that coordinate. The wet part of the sheet is at $s<D$ and is subjected to the Laplace pressure 
\begin{equation}
\label{laplace-pressure}
p=\gamma/R_d.
\end{equation}
In the dry part, on the other hand, the pressure is atmospheric: $p=0$.
Given $\kappa$, the local angle with respect to the horizontal direction is found by
\begin{equation}
\partial_s\theta=\kappa, \label{eq:theta1}
\end{equation}
and the Cartesian coordinates along the sheet satisfy
\begin{align}
\partial_s x&=\cos \theta,
&\partial_s y&=\sin\theta.
\label{eq:xy1}
\end{align}
Instead of $n_\parallel$ and $n_\perp$, one may use the Cartesian components $n_x= n_\parallel\cos\theta-n_\perp\sin\theta$ and $n_y=n_\parallel\sin\theta+n_\perp \cos\theta$. This yields
\begin{subequations}
\label{ODE-general-p-wet}
\begin{align}
\label{ODE-general-p-wet-B}
B\, \partial_s^2\theta&=n_x\, \sin\theta-n_y\, \cos\theta, \\
\label{ODE-general-p-wet-nx-ny}
\partial_s n_x&=-p\, \sin\theta,\quad \partial_s n_y=p\, \cos\theta.
\end{align}
\end{subequations}
Note that the above equations can also be derived through energy minimisation, as detailed in Ref.~\cite{Neukirch13}. In the dry region ($s>D$), we have $(n_x,n_y)=(T,0)$ everywhere. Hence, multiplying Eq.~(\ref{ODE-general-p-wet-B}) by $\partial_s\theta$ and integrating, we obtain, for an infinite domain with $\lim_{s\to\infty}(\theta,\kappa)=(0,0)$,
\begin{equation}
\kappa_D=-2\left(T/B\right)^{1/2}\sin\left( \theta_D/2\right), \label{BC:kappaD}
\end{equation}
where $\kappa_D$ and $\theta_D$ respectively denote the curvature and angle at $s=D$. On the wet side of this point, the force balance is
\begin{equation}
n_x(D)=T-\gamma\, \cos\beta, \quad n_y(D)=\gamma\, \sin\beta. \label{BC:nxy}
\end{equation}
Assuming symmetric shapes and imposing YLD law at the contact line, we have,
\begin{equation}
\theta(0)=0, \quad \theta_D+\beta=\theta_Y.
\label{BC:theta(D)}
\end{equation}
Next, translation invariance allows us to fix the values 
\begin{equation}
\label{x-y-0}
x(0)=0, \quad y(0)=0.
\end{equation}
Finally, the geometrical constraints (\ref{nobending:eq:A}) and (\ref{geometry-zero-B-xd}) become
\begin{subequations}
\label{BC:A,xD}
\begin{align}
\label{BC:A}
A&=R_d^2\left(\beta-\frac{\sin2\beta}{2}\right)+2\int_0^Dx(s)\sin\theta(s)\,\mathrm{d} s, \\
\label{BC:xD}
x_D&=R_d\, \sin\beta.
\end{align}
\end{subequations}

In the wet region $0\leq s<D$, we have to solve Eqs.~(\ref{eq:nperp1})-(\ref{eq:xy1}) with the boundary conditions and global constraint in Eqs.~(\ref{BC:kappaD})-(\ref{BC:A,xD}). Note that there are 9 conditions because there are 6 differential equations and 3 unknown parameters, $\beta$, $\theta_D$, and $R_d$. Notice that, integrating the last of Eqs.~(\ref{ODE-general-p-wet-nx-ny}) between $0$ and $D$ and using the first of Eqs.~(\ref{eq:xy1}) and (\ref{x-y-0}) together with the second of Eqs.~(\ref{BC:nxy}) and Eqs.~(\ref{BC:xD}), we obtain
\begin{equation}
\label{ny0}
n_y(0)=0.
\end{equation}
Finally, by virtue of the geometrical constraint (\ref{BC:xD}) and Eq.~(\ref{laplace-pressure}), the capillary pressure can also be expressed as
\begin{equation}
p= \gamma \,x_D^{-1}\sin\beta.
\label{gammasinbeta/xD}
\end{equation}

\subsection{Vesicle state}

In the vesicle state, the sheet is in self-contact, see Fig.~\ref{fig04-setup}(b). At the point of self-contact, $s=\ell$, there is a localised reaction force, $F_c$, which modifies the second of Eqs.~(\ref{eq:nperp1}) and the first of Eqs.~(\ref{ODE-general-p-wet-nx-ny}) as
\begin{subequations}
\label{eq:nperp1bis}
\begin{align}
\label{eq:nperp1bis-nperp}
\partial_s n_\perp &= p-\kappa\, n_\parallel +F_c \, \delta(s-\ell), \\
\label{eq:nperp1bis-nx}
\partial_s n_x &= -p\, \sin \theta - F_c\, \delta(s-\ell).
\end{align}
\end{subequations}
Equations (\ref{eq:nperp1bis}) brings two new unknown parameters, $F_c$ and $\ell$, into the problem which are fixed by two new boundary conditions:
\begin{align}
x_\ell&=0,
&\theta_\ell&=\pi/2,
\label{BC:xl}
\end{align}
where subscript $\ell$ means evaluation at $s=\ell$. While the total area $A$ of the fluid is still given by Eq.~(\ref{BC:A}), it is now split in two parts, $A_1$ and $A_2$, respectively below and above the contact point. If the contact is such that no fluid is allowed through, both $A_1$ and $A_2$ are in principle constrained to a fixed value, instead of just $A$. In response to this new constraint, the pressure $p$ differs from the capillary pressure inside the vesicle. A complete theory should therefore discuss the solution not only as a function of $A$, but also of $A_1$.  However, there is no general rule that governs how $A$ should split between $A_1$ and $A_2$ and, hence, what the vesicle pressure should be. In the absence of a law dictating the ratio $A_1/A_2$, we will assume that it is  free to vary with only the constraint $A_1+A_2=A$   and $p_1=p_2$.

\subsection{First integrals and alternative independent variables}

In the wet part of the elastic sheet, combining Eqs.~(\ref{eq:xy1}) with Eqs.~(\ref{ODE-general-p-wet}) and the first and last of Eqs.~(\ref{eq:nperp1}), we obtain
\begin{subequations}
\label{eq-x-y-kappa}
\begin{align}
\label{eq-x-y}
&\partial_s\left(x-n_y/p\right)=0, \quad \partial_s\left(y+n_x/p\right)=0,\\
\label{eq-kappa}
&\partial_s\left(B\kappa^2/2+n_\parallel\right)=0.
\end{align}
\end{subequations}
This allows us, on the one hand, to deduce the shape of the wet part of the sheet once the tractions and $p$ are known and, on the other hand, to write
\begin{equation}
\frac{B}{2}\, \kappa^2+n_\parallel=H,
\label{eq:defH}
\end{equation}
where $H$ is constant. Evaluating Eq.~(\ref{eq:defH}) at $s=D$ and using the boundary conditions~(\ref{BC:kappaD}) and (\ref{BC:nxy}) together with the relation between $n_\parallel$ and $n_x$ and $n_y$, one finds
\begin{equation}
H=T-\gamma\cos\theta_Y.
\label{eq:defH2}
\end{equation}
In the case of self-contact, at $s=\ell<D$, we have $\theta(\ell) = \pi/2$, so that  $n_\parallel(\ell) = n_y(\ell)$. Next, integrating the first of Eqs.~(\ref{eq-x-y}) between $0$ and $\ell$ and using $x(0)=x(\ell)=n_y(0)=0$, we get $n_\parallel(\ell)=0$. Therefore, evaluating Eq.~(\ref{eq:defH}) at $s=\ell$ leads to
\begin{equation}
H=T-\gamma\cos\theta_Y = \frac{B}{2}\, \kappa_\ell^2,
\label{eq:defH3}
\end{equation}
so that the curvature at the contact point in the vesicle state vanishes when $T=T^{\star} = \gamma \cos\theta_Y$. Since $\kappa_\ell^2\ge 0$, the vesicle state exists only for $T\ge T^{\star}$.

Using Eq.~(\ref{eq:defH}) to eliminate $n_\parallel$, the second of Eqs.~(\ref{eq:nperp1}) becomes
\begin{equation}
\partial_s n_\perp=p-H\, \kappa+ \frac{B}{2}\, \kappa^3.
\label{eq:nperp2}
\end{equation}

While $s$ appears as the most natural variable to express all the physical quantities along the sheet, one should note that other variables can be more advantageous. In particular, since the differential system is autonomous in $s$, one can reduce its order by using one of the dynamical variables as the independent one and seeking all the others quantities as functions of it. One useful choice is $\theta$. Let us introduce
\begin{equation}
\label{rel-kappa-q}
\kappa^2=2q(\theta(s)).
\end{equation}
The variable $q$ may be regarded as a measure of the density of bending energy. By differentiating each side of Eq.~(\ref{rel-kappa-q}) with respect to $s$, one finds that $\partial_s\kappa=\partial_\theta q$. Hence, the first of Eqs.~(\ref{eq:nperp1}) and Eq.~(\ref{eq:nperp2}) become
\begin{align}
\label{temp}
B\, \partial_\theta q&=-n_\perp, 
&\partial_\theta n_\perp&=\frac{p}{\kappa}- H+B\, q.
\end{align}
Differentiating the first equation above with respect to $\theta$, we thus obtain
\begin{equation}
B\left(\partial^2_{\theta}q+q\right)+\frac{p}{\kappa}= H.
\label{eq:q1}
\end{equation}
This last formulation leads to considerable simplification when either $p/\kappa$ or $B(\partial^2_{\theta}q+q)$ dominates the left hand side. 

Another useful trick is to treat $\kappa$ as the independent variable. Indeed, using $\partial_s n_{\bot} = (\partial_{\kappa} n_{\bot}) \partial_s \kappa$ and the first of Eqs.~(\ref{eq:nperp1}), Eq.~(\ref{eq:nperp2}) becomes
\begin{equation}
\partial_\kappa n_\perp^2= 2 B\left(-p+ H\, \kappa -\frac{B}{2}\kappa^3\right),
\label{eq:nperpkappa}
\end{equation}
which is simple to integrate. Together with Eq.~(\ref{eq:defH}), this equation yields the tractions, and hence $x$ and $y$, directly as functions of $\kappa$.

Note that a rather complete treatment of Eqs.~(\ref{eq:nperp1}) in terms of Jacobi functions and elliptic integrals of the first and third kinds was developed in Ref.~\cite{Djondjorov2011}. While the approach followed  in Sec.~\ref{sec:partialwrap}  is only asymptotically exact, as $\Gamma\to\infty$, it has the advantage of involving mostly elementary functions, hence expressions that are easier to interpret (see also the comment at the end of Sec.~\ref{shape-t=ts}).

\begin{figure*}[!t]
\centering
\includegraphics[width=\textwidth]{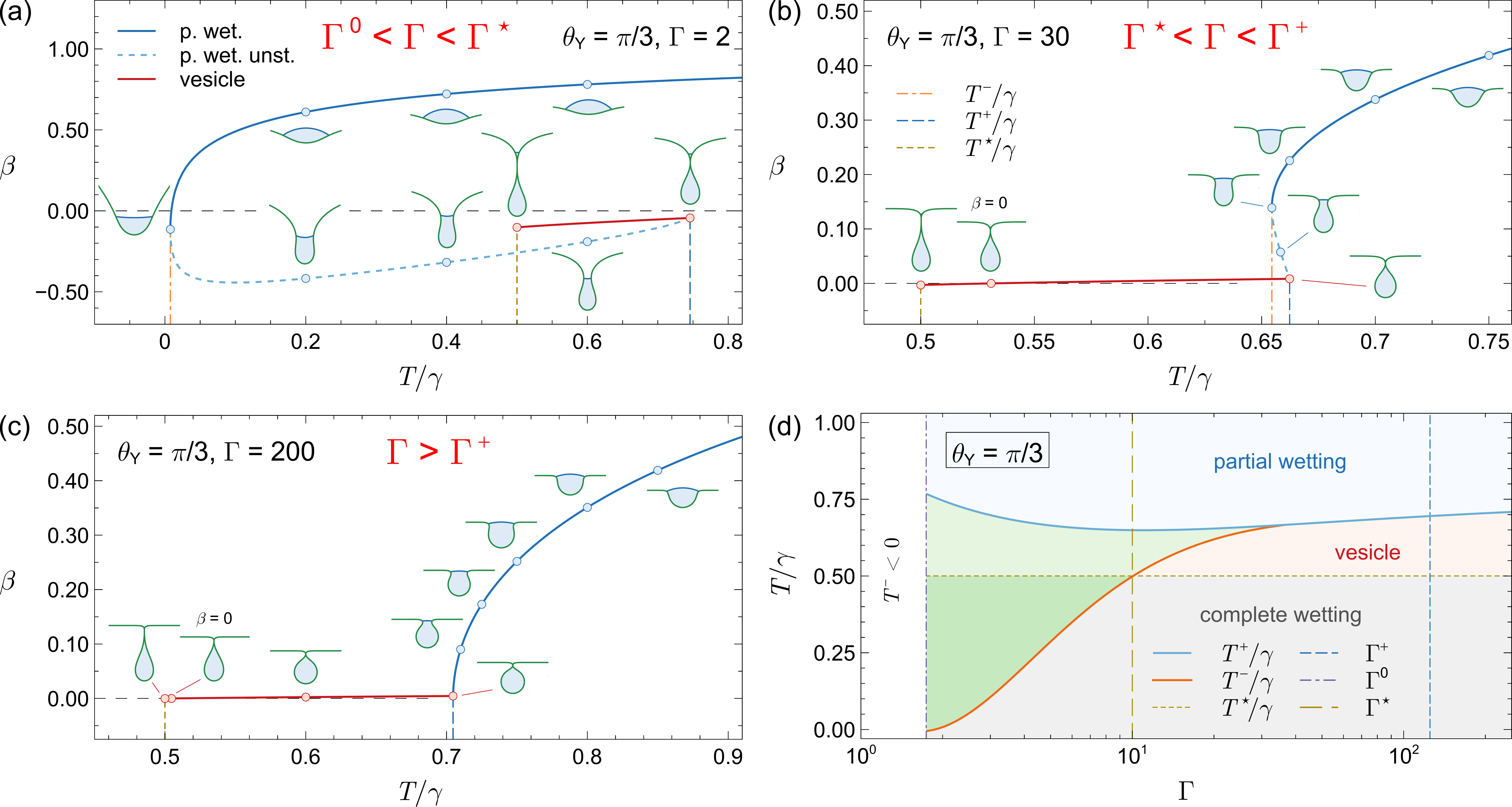}
\caption{(a)-(c) Bifurcation diagrams showing the evolution of $\beta$ as a function of $T/\gamma$ for $\theta_Y=\pi/3$ and three values of $\Gamma = \gamma\, A/B$. Insets: representative shapes of the system. (a) For $\Gamma=2$, the system bifurcates at $T =T^{-}$ from partial wetting to complete wetting. (b) For $\Gamma=30$, the transition between partial wetting and vesicle is subcritical with region of bistability, $T^{-}< T <T^{+}$, where both states exist. (c) For $\Gamma=200$, the bifurcation between partial wetting and vesicle is supercritical with a continuous transition as the applied tension varies. For both (b) and (c), complete wetting occurs when $T <T^{\star}$. (d) Evolution of $T^{-}$ and $T^{+}$ as a function of $\Gamma$ for $\theta_Y=\pi/3$. At $\Gamma = \Gamma^0$, $\Gamma = \Gamma^{\star}$ and $\Gamma = \Gamma^+$, the tension $T^{-}$ is vanishing, equal to $T^{\star}=\gamma \cos \theta_Y$ and equal to $T^{+}$, respectively.
}
\label{fig-bifurcation-diag-numerics}
\end{figure*}

\subsection{Solution in the dry region}
\label{sec:general-eq-dry-region}

In the dry part of the sheet, where $n_x=T$ and $n_y=0$, Eq.~(\ref{ODE-general-p-wet}) reduces to
\begin{equation}
\label{eq-dry}
B\partial_s^2\theta-T\sin\theta=0,
\end{equation}
with boundary conditions $\theta(D)=\theta_D$ and $\theta(L)=0$. Letting $L\to\infty$, the solution has the exact form
\begin{equation}
\theta(s)=4\arctan\left[\tan\left[\frac{\theta_D}4\right]\exp\left[-\sqrt{\frac{T}{B}}\left(s-D\right)\right]\right],
\label{thetadry}
\end{equation}
whose derivative at $s=D$ is given by Eq.~(\ref{BC:kappaD}). On the other hand, multiplying Eq.~(\ref{eq-dry}) by $\kappa=\partial_s \theta$ and integrating we have $\kappa=-2\left(T/B\right)^{1/2}\sin\left( \theta/2\right)$. Parametrizing $x$ and $y$ with $\theta$ and using this expression, Eqs.~(\ref{eq:xy1}) become
\begin{align}
\partial_\theta x&=-\sqrt{\frac{B}{T}}\frac{\cos\theta}{2\sin\left(\theta/2\right)},
&\partial_\theta y&=-\sqrt{\frac{B}{T}}\frac{\sin\theta}{2\sin\left(\theta/2\right)}.
\end{align} 
This yields
\begin{subequations}
\begin{align}
x&=c_x-\left(B/T\right)^{1/2}\left[2\cos\left(\theta/2\right)+\ln\left(\tan\left(\theta/4\right)\right)\right], \\
y&=c_y-2\left(B/T\right)^{1/2}\sin\left(\theta/2\right),
\end{align}
\end{subequations}
where $c_{x,y}$ are constants of integration. This illustrates how all the variables can be expressed in terms of $\theta$. If we substitute $\theta$ by the right hand side of Eq.~(\ref{thetadry}), we obtain their explicit dependence on $s$.

\section{Numerical study}
\label{sec:num}

As shown in Sec.~\ref{sec:general-eq-dry-region}, the shape of the sheet in the dry region, $D < s\le L$, is known explicitly in the limit of a long sheet, $L\gg \sqrt{A}$. To solve the problem in the wet region, $0\le s \le D$, for the partial wetting state, we need to integrate numerically Eqs.~(\ref{ODE-general-p-wet}) with the associated boundary conditions (\ref{BC:kappaD})-(\ref{BC:theta(D)}) and geometric constraints (\ref{BC:A,xD}). For the vesicle state, the first of Eqs.~(\ref{ODE-general-p-wet-nx-ny}) is replaced by Eq.~(\ref{eq:nperp1bis-nx}) and we must consider two additional boundary conditions given by Eqs.~(\ref{BC:xl}). For this purpose, a shooting method is used where the boundary value problem is transformed into an initial value problem and the unknown initial conditions are varied until the boundary conditions are satisfied. In this way, we may simulate the system for values of $\Gamma = \gamma\, A/B$ ranging from 0 to about 300.

Whatever the values of $\Gamma$ and $0<\theta_Y \le \pi/2$, the system is always in a partial wetting state when the applied tension $T$ is large enough. Indeed, in the limit $T\to \infty$, the sheet is flat and behaves as an undeformable substrate. In this case, complete wetting is possible only if $\theta_Y = 0$.

When the applied tension decreases, transitions towards the vesicle state and complete wetting occur. Figure~\ref{fig-bifurcation-diag-numerics}(a)-(c) shows bifurcation diagrams for $\theta_Y = \pi/3$ and three representative values of $\Gamma$ where $\beta$ is used as an order parameter and the applied tension as the bifurcation parameter. These plots highlight the existence of three distinctive values of the applied tension. 

Similarly to the case of a zero bending modulus discussed in Sec.~\ref{sec:zero-B}, there are $T^{\star}=\gamma\cos \theta_Y$ and $T^{+}(\theta_Y,\Gamma)$ delimiting the domain of existence of the vesicle state. At $T=T^{\star}$, the curvature at the contact point, $s=\ell$, vanishes (see Eq.~(\ref{eq:defH3}) and Fig.~\ref{fig-shapes}(c)). This value of the tension is thus the smallest one for which a vesicle state exists. At $T=T^{+}(\theta_Y,\Gamma)$, the self-contact occurs with a vanishing contact force, i.e. $F_c=0$ (see Fig.~\ref{fig-shapes}(b)). Beyond this applied tension, there is no longer self-contact. 

In addition, for $\Gamma<\Gamma^{+}(\theta_Y)$, we find that a new special value of the tension, $T^{-}(\theta_Y,\Gamma)$, shows up for a finite bending modulus. This is the smallest tension for which partial wetting states exist. As $T \to T^{-}(\theta_Y,\Gamma)$, the curve $\beta(T)$ develops a vertical slope [Figs.~\ref{fig-bifurcation-diag-numerics}(a,b)] but there is no self-contact in contrast to the case of an infinitely bendable sheet. In addition to $\Gamma^{+}(\theta_Y)$, the bifurcation diagrams highlight the existence of two other special values of the parameter $\Gamma$, that we denote $\Gamma^0(\theta_Y)$ and $\Gamma^{\star}(\theta_Y)$. These three values are marked by the three vertical dashed lines in Fig.~\ref{fig-bifurcation-diag-numerics}(d). 

When $\Gamma < \Gamma^0(\theta_Y)$, $T^{-}(\theta_Y,\Gamma)$ is negative, such that when the tensile load $T$ is reduced, the system remains in a partial wetting state down to $T=0$.

When $\Gamma^0(\theta_Y) <\Gamma < \Gamma^{\star}(\theta_Y)$, as in Fig.~\ref{fig-bifurcation-diag-numerics}(a), the system bifurcates from partial wetting to complete wetting as the decreased applied tension reaches $T^{-}(\theta_Y,\Gamma)$. In this case, the vesicle branch is not reached by decreasing the applied tension from $T\gg \gamma$. 

When $\Gamma^{\star}(\theta_Y)< \Gamma < \Gamma^{+}(\theta_Y)$, as in Fig.~\ref{fig-bifurcation-diag-numerics}(b), the transition between vesicle and partial wetting is subcritical and there is a region of applied tension where both states coexist. There are thus discontinuous transitions between both states at $T=T^{-}(\theta_Y,\Gamma)$ and $T=T^{+}(\theta_Y,\Gamma)$. In this case, the partial wetting branch that bifurcates subcritically from the vesicle branch at $T^{+}$ [blue dashed line in Fig.~\ref{fig-bifurcation-diag-numerics}(b)] is unstable; it only becomes stable at the limit point $T^{-}$. For tensions in the range $T^{-}<T<T^{+}$, three values of $\beta$ are possible, each corresponding to a distinct steady state. The middle one, belonging to the blue dashed branch in Fig.~\ref{fig-bifurcation-diag-numerics}(b) yields a local maximum of the energy and is therefore unstable.

When $\Gamma > \Gamma^{+}(\theta_Y)$, as in Fig.~\ref{fig-bifurcation-diag-numerics}(c), the bifurcation is supercritical with a continuous transition between both states. The transition occurs at $T=T^{+}(\theta_Y,\Gamma)$, at which value the partial wetting state is stable and there is self-contact with $F_c=0$.

\begin{figure}[!t]
\centering
\includegraphics[width=\columnwidth]{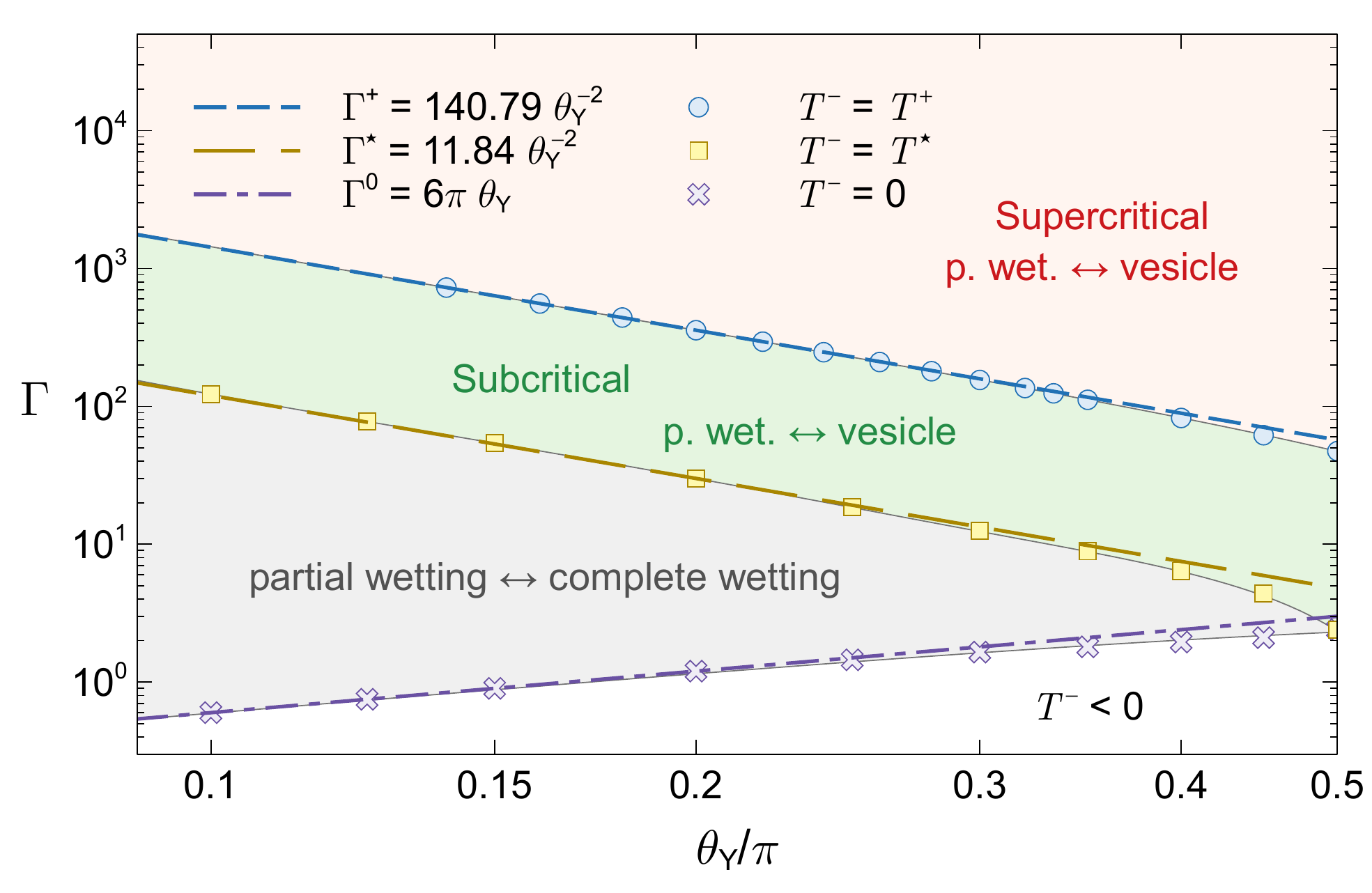}
\caption{Evolution of $\Gamma^0$, $\Gamma^{\star}$ and $\Gamma^{+}$ as a function of $\theta_Y$ delimiting the possible types of bifurcation in the $(\theta_Y,\Gamma)$ space. The grey, green, and pink regions correspond to bifurcation diagrams depicted in Fig.~\ref{fig-bifurcation-diag-numerics} (a), (b), and (c), respectively.}
\label{fig-sub-supercritical}
\end{figure}

This complex phase diagram is summarized in Fig.~\ref{fig-bifurcation-diag-numerics}(d), which shows that, as $\Gamma$ increases, the difference between $T^{+}$ and $T^{-}$ decreases. When $\Gamma^0(\theta_Y) <\Gamma < \Gamma^{\star}(\theta_Y)$, $T^{-}$ is smaller than $T^{\star}$ and the bifurcation diagram is similar to the one shown in Fig.~\ref{fig-bifurcation-diag-numerics}(a) in this region. When $\Gamma^{\star}(\theta_Y)< \Gamma < \Gamma^{+}(\theta_Y)$, $T^{-}$ is larger than $T^{\star}$ while still smaller than $T^{+}$. The bifurcation diagram in this region is similar to the one shown in Fig.~\ref{fig-bifurcation-diag-numerics}(b). When $\Gamma = \Gamma^{+}(\theta_Y)$, we have the equality $T^{-}=T^{+}$. Finally, when $\Gamma > \Gamma^{+}(\theta_Y)$, partial wetting states, i.e. solutions of Eqs.~(\ref{ODE-general-p-wet}), display self-crossing for $T<T^{+}$ and must therefore be discarded. Hence, the system is in a vesicle state when $T^{\star}< T<T^{+}$. This corresponds to the bifurcation diagram shown in Fig.~\ref{fig-bifurcation-diag-numerics}(c). 

The algorithm to compute $\Gamma^{+}(\theta_Y)$ is described in Appendix~\ref{app:gammac}. The result of this computation shows that when $\theta_Y$ is small enough, $\Gamma^{+}(\theta_Y)\sim \theta_Y^{-2}$ (see Fig.~\ref{fig-sub-supercritical}). Therefore, whatever the value of $\theta_Y$ is, there always exist values of $\Gamma$ such that the transition is supercritical. However, this shows that the limit $\Gamma \to \infty$ together with $\theta_Y \to 0$ is subtle and will not be considered in the asymptotic theory presented in Sec.~\ref{sec:partialwrap}. Specifically, we will assume $\Gamma \gg \Gamma^{+}(\theta_Y)$ with $\theta_Y=O(1)$.

It is also possible to compute $\Gamma^{0}(\theta_Y)$ for which $T^{-}=0$ and $\Gamma^{\star}(\theta_Y)$ for which $T^{-}=T^{\star}$. For this purpose, $T^{-}$ is obtained for given $\theta_Y$ and $\Gamma$ and the latter is varied by small increments. For each value of $\Gamma$, $T^{-}$ is computed until it reaches $0$ or $T^{\star}$. The result of this computation is shown in Fig.~\ref{fig-sub-supercritical}. When $\theta_Y$ is small enough, $\Gamma^{0} \sim \theta_Y$ and $\Gamma^{\star} \sim \theta_Y^{-2}$.

More details on the various T-dependent quantities are described in Appendix~\ref{app-num}.
  

The theory shows that the vesicle state exists only when the tension is larger than $T^{\star}$, see Eq.~(\ref{eq:defH3}). The numerical results show that this state exists only when the tension is smaller than $T^{+}(\theta_Y,\Gamma)$, which tends to $\gamma\cos^2(\theta_Y/2)$ as $\Gamma\to\infty$, in agreement with the limit of vanishing bending modulus discussed in Sec.~\ref{sec:zero-B}, see Eq.~(\ref{Tnb-def}). For $\Gamma=\infty$, the shape of the vesicle is predicted to be circular with radius $(A/\pi)^{1/2}$ independently of tension. Numerical results, on the other hand, show that the vesicle shape can significantly depart from a circle, as can be seen in Fig.~\ref{fig-bifurcation-diag-numerics}(b) for $\Gamma=30$. In the vicinity of $T=T^{\star}$, the vesicle has a teardrop shape. The range of tensions for which the vesicle is markedly non-circular shrinks as $\Gamma\to\infty$ but nevertheless remains significant even for $\Gamma=200$; this is explained by the asymptotic theory of Sec.~\ref{sec:partialwrap}.

A non-vanishing bending modulus has thus a significant impact on the vesicle shape, and not merely a boundary layer near the point of self-contact. The vesicle shape is controlled by two length scales. The radius of curvature of the sheet away from the contact point scales like the size of the drop, $\sqrt{A}$. However, the radius of curvature at the contact point does not scale like $\ell_{\text{BC}}$ as one could expect. Instead, it scales as
\begin{equation}
\lambda_{\text{BC}}\equiv \sqrt{\frac{B}{2H}}= \sqrt{\frac{B}{2\left(T-\gamma\cos\theta_Y\right)}},
\label{eq:def:lambdaBC}
\end{equation}
 as follows from Eq.~(\ref{eq:defH3}). For a fixed value of $T>T^{*}$, it tends to zero as $B\to0$ or, equivalently, as $\Gamma\to\infty$. The shape adopted by the vesicle is thus essentially controlled by the length $\sqrt{A}$ and tends to a circle, except in a boundary layer near the contact point. By contrast, when $T \to T^{\star}$ but $B$, or $\Gamma$, is kept constant, $\lambda_{\text{BC}}$ diverges. In that limit, the vesicle shape necessarily departs from a circle and this is what we investigate next.

\section{Asymptotics of the Vesicle Solution}\label{sec:partialwrap}

We now analyse the solution depicted in Fig.~\ref{fig04-setup}\,(b). We make the assumption that self-contact at $s=\ell$ takes place in the wet part of the sheet, i.e. $\ell \le D$. This is numerically verified for $\theta_Y<\pi/2$. The solution for the dry part of the sheet is already known, see Sec.~\ref{sec:general-eq-dry-region}. The asymptotic solution for the wet part of the sheet is derived in Secs.~\ref{ves-beyond-contact} ($\ell <s<D$) and \ref{ves-before-contact} ($s<\ell$).

\subsection{Above the contact point}
\label{ves-beyond-contact}

In the range $\ell <s<D$, one has $\kappa=O(x_D^{-1})$ and numerical solutions indicate that $\beta$ is small, so that $p=\gamma/R_d=\gamma \sin\beta/x_D\simeq \gamma \beta/x_D$ (see Eq.~(\ref{gammasinbeta/xD})). Using Eq.~(\ref{rel-kappa-q}), we thus have
\begin{align}
\frac{p/\kappa}{Bq}&=O\left(\frac{\gamma\, \beta\, x_D^2}{B}\right)\ll1
&\text{if}&
&\beta&\ll \frac{A}{\Gamma\, x_D^2} .
\label{smallness}
\end{align}
Under this hypothesis, we may neglect $p/\kappa$ in Eq.~(\ref{eq:q1}) and obtain
\begin{equation}
\frac{B}{2} \kappa^2 \equiv Bq\simeq H\left[1+d\cos\left(\theta+\psi\right)\right],
\label{eq:q2}
\end{equation}
where $d$ and $\psi$ are constants of integration and where we have used the relation (\ref{rel-kappa-q}) between $q$ and $\kappa$. Evaluating Eq.~(\ref{eq:q2}) at the contact point $s= \ell$, where $\theta=\pi/2$, and comparing with Eq.~(\ref{eq:defH3}), we find $\psi=0$. Next, using the boundary condition (\ref{BC:kappaD}) and the relation~(\ref{eq:defH2}) between $H$ and $T$, the evaluation of Eq.~(\ref{eq:q2}) at $s=D$ leads to
\begin{equation}
\label{d-def}
d=\frac{\gamma\cos\theta_Y\left(1-\cos\theta_D\right)}{H \cos\theta_D}-1.
\end{equation}
Knowing that $\text{sgn}(\kappa)<0$ in the region $\ell <s<D$, the curvature is readily obtained from Eq.~(\ref{eq:q2}):
\begin{equation}
\kappa(\theta) \simeq -\lambda_\text{BC}^{-1}\sqrt{1+d\cos\theta},
\end{equation}
where $\lambda_\text{BC}$ is given by Eq.~(\ref{eq:def:lambdaBC}). Hence, with $\partial_\theta x=\cos\theta/\kappa$, we obtain
\begin{equation}
x_D =\lambda_\text{BC}\int_{\theta_D}^{\pi/2}\frac{\cos\theta\,\mathrm{d} \theta}{\sqrt{1+d\cos\theta}}.
\end{equation}
From this expression, the asymptotic limit (\ref{smallness}) under which the present derivation holds is simply $\beta\ll1$. Hence $\theta_D\sim\theta_Y$ [see Eq.~(\ref{BC:theta(D)})] yielding $d \simeq (\gamma/H) \left(1-\cos\theta_Y\right)-1$ and thus
\begin{equation}
x_D \simeq \lambda_\text{BC}\sqrt{\frac{H}{\gamma}}\int_{\theta_Y}^{\pi/2}\frac{\cos\theta\,\mathrm{d} \theta}{\sqrt{H/\gamma+( 1-T/\gamma)\cos\theta}}.
\end{equation}
On the other hand, the equation $\partial_\theta y=\sin\theta/\kappa$ yields, with Eq.~(\ref{eq:q2}),
\begin{equation}
\label{ves-y}
y(\theta)\simeq y_D+\frac{B}{Hd}[\kappa_D-\kappa(\theta)].
\end{equation}
The area $A_2$ [see Fig.~\ref{fig04-setup}(b)] is given by
\begin{equation}
A_2 =R_d^2\left(\beta-\frac12\sin2\beta\right)+x_D y_D - 2\int_{x_{\ell}}^{x_D}y\,\mathrm{d} x.
\end{equation}
Using $\mathrm{d} x = \partial_{\theta} x \, \mathrm{d}\theta = \cos \theta \, \mathrm{d}\theta/\kappa$ and Eq.~(\ref{ves-y}), the integral can be computed to obtain
\begin{equation}
A_2 = x_D^2\left(\frac{2\beta-\sin2\beta}{2\sin^2\beta}\right)+\frac{2B}{Hd}\left(\sin\theta_D-1-\kappa_D x_D\right).
\end{equation}
It follows that $A_2/A=O(\Gamma^{-1})$ and thus vanishes in the limit $\Gamma \to \infty$ in agreement with the zero-bending case discussed in Sec.~\ref{sec:zero-B}. 

For future reference, let us finally note that, using Eq.~(\ref{temp}), we have 
\begin{equation}
\lim_{s\to\ell^+} n_\perp=\lim_{\theta\to\pi/2}\left(-B\partial_\theta q\right)=\gamma\left(\frac{\cos\theta_Y}{\cos\theta_D}-T\right).
\label{eq:nperp2bis}
\end{equation}

\begin{figure}
\centering
\includegraphics[width=\columnwidth]{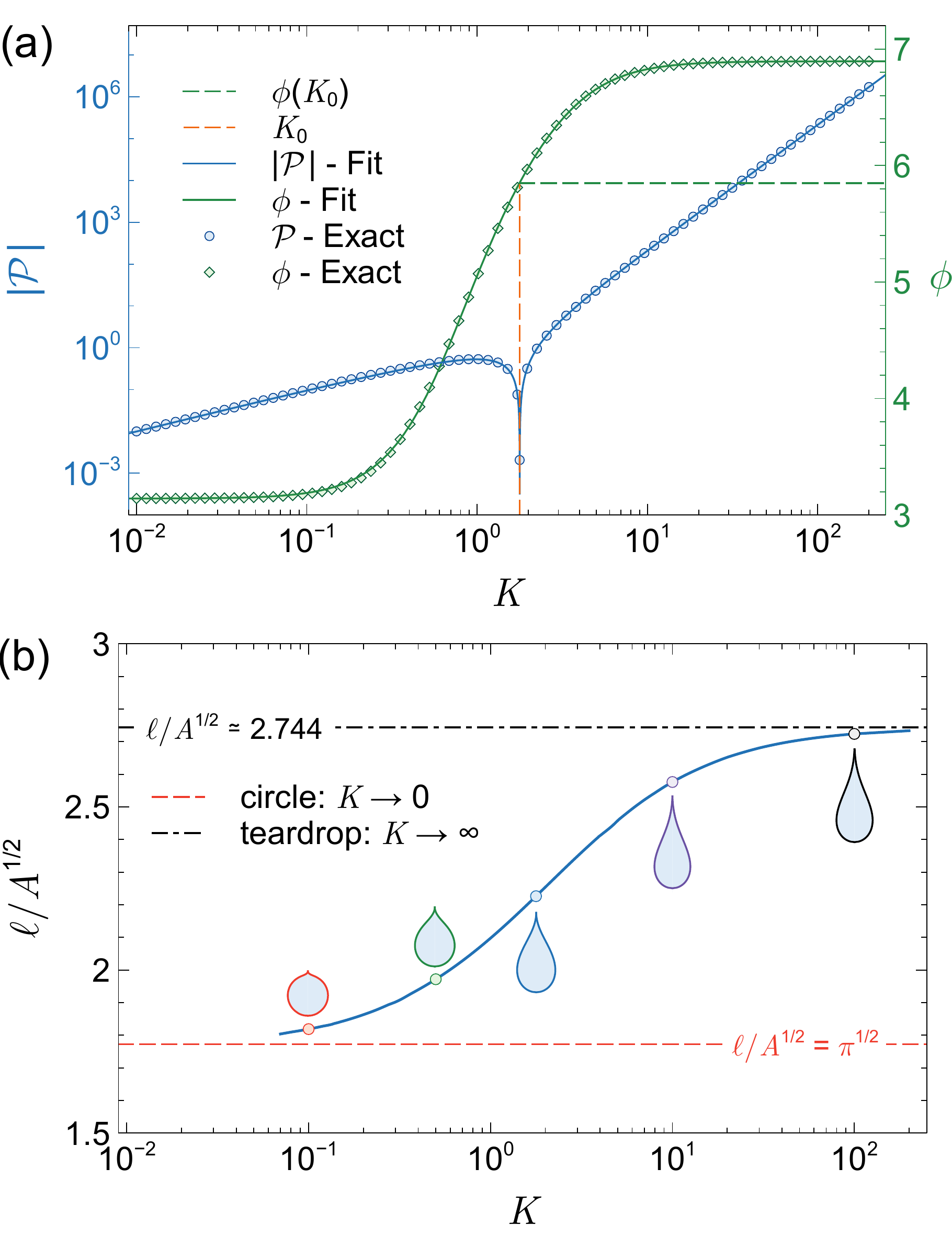}
\caption{(a) Evolution of the functions $|\mathcal{P}|$ and $\phi$, defined by Eqs.~(\ref{eq:P(K0)}) and (\ref{eq:phi(K0)}), as a function of $K$ together with the two approximations (\ref{fit:PP}) and (\ref{fit:phi}). $\mathcal{P}$ vanishes at $K=0$ and at $K=K_0 \simeq 1.77842$ where $\phi(K_0)\simeq 5.84946$. It has the asymptotic behaviours $\mathcal{P}\simeq K$ and $\mathcal{P}\simeq -3K^3/14$ when $K\to0$ and $K\to\infty$, respectively. As $K\to0$, the vesicle becomes a circle, so that $\phi(0)=\pi$. (b) Evolution of (half) the length of the vesicle, $\ell$, as a function of $K$. Some representative vesicle shapes are shown along the curve. As $K\to0$, the vesicle shape approaches a circular shape of radius $(A/\pi)^{1/2}$ whereas, as $K\to\infty$, the vesicle shape tends to a teardrop.}\label{fig:P(K0)}
\end{figure}

\subsection{Below the contact point}\label{sec:Kless392}
\label{ves-before-contact}

Inside the vesicle, i.e. $s<\ell$, our numerical solutions indicate that all terms in Eq.~(\ref{eq:q1}) are generally of the same order. In this case, the alternative Eq.~ (\ref{eq:nperpkappa}) is more convenient to analyze. The curvature decreases from a value
\begin{subequations} 
\begin{align}
\kappa_0&>0 &\text{at}\quad &s=0, \\
\text{to} \quad \kappa_\ell&=-\lambda_\text{BC}^{-1}<0 &\text{at}\quad  &s=\ell.
\end{align}
\end{subequations} 
From Eq.~(\ref{eq:nperpkappa}), we directly get
\begin{equation}
 n_\perp^2 =2Bp\left[\kappa_0-\kappa\right]-BH \left[\kappa_0^2-\kappa^2\right] +\frac{B^2}{4}\left[\kappa_0^4-\kappa^4\right],
 \label{sol:n1}
\end{equation}
where we used the fact that $n_\perp=\partial_s\kappa=0$ at $s=0$ [see Eqs.~(\ref{ny0}, (\ref{ODE-general-p-wet-B}), and (\ref{eq:nperp1})]. 
Let us rescale $\kappa$, $p$, and introduce the parameter $K$ as follows
\begin{align}
k&=\frac{\kappa}{\kappa_0},
&p&=\sqrt{\frac{2H^3}{B}}\mathcal{P},
&K&= \kappa_0 \sqrt{\frac{B}{2H}}=\kappa_0\lambda_\text{BC}.
\label{def:k,PP,K0}
\end{align}
With these new notations, we may rewrite Eq.~(\ref{sol:n1}) as
\begin{subequations}
\label{sol:n2-def:NN}
\begin{align}
\label{sol:n2}
 n_\perp &=B\kappa_0^2 \; \mathcal{N}_\perp\left(k;\mathcal{P},K\right), \\
\label{def:NN}
\mathcal{N}_\perp& =\sqrt{1-k}\left[\frac{\mathcal{P}}{K^3}+\frac{1+k}{4}\left(1+k^2-\frac{2}{K^2}\right)\right]^{1/2}.
\end{align}
\end{subequations}
Next, using the first of Eqs.~(\ref{eq:nperp1}), we have $\kappa=\partial_s\theta=\partial_s\kappa \, \partial_\kappa\theta=-\left(B\kappa_0\right)^{-1} n_\perp \partial_k\theta$. Hence, using this last relation and Eq.~(\ref{sol:n2}), we obtain
\begin{equation}
\label{def-theta}
\theta(k;K)=\int_{k}^1 \frac{k'}{\mathcal{N}_\perp\left(k';\mathcal{P},K\right)}\,\mathrm{d} k',
\end{equation}
where we used the fact that $\theta$ vanishes when $k=1$, i.e. when $s=0$. Evaluating this expression at the contact point, where $k=-1/K$, yields a condition on $\mathcal{P}$:
\begin{equation}
\int_{-1/K}^1 \frac{k}{\mathcal{N}_\perp\left(k;\mathcal{P},K\right)}\,\mathrm{d} k=\frac{\pi}{2}.
\label{eq:P(K0)}
\end{equation}
The solution of this equation is universal and denoted by $\mathcal{P}=\mathcal{P}(K)$, see Fig.~\ref{fig:P(K0)}(a). Note in particular that it doesn't depend on $\Gamma$. The formula above implies that the curvature varies monotonously from $\kappa_0$ to $\kappa_\ell$. This is only true up to $K\simeq 3.9207$. For larger values, $\mathcal{N}_\perp$ necessarily passes by zero and the integral in the left hand side of Eq.~(\ref{eq:P(K0)}) must be split into two parts. We ignore this difficulty in this section and give details in Appendix~\ref{app:K392}. A good numerical fit of $\mathcal{P}(K)$, valid for all $K$, is given by
\begin{equation}
\mathcal{P}(K)\approx \left(K_0-K\right)\left(\frac{K}{K_0}+\frac{3}{14}\frac{K^3}{K+0.846}\right),
\label{fit:PP}
\end{equation}
where $K_0=1.77842$ [Fig.~\ref{fig:P(K0)}(a)]. The value $K_0$ is such that $\mathcal{P}$, and hence $\beta$, vanishes.

Once the function $\theta(k;K)$ is known, the equation for $x$ can be rewritten as
\begin{equation}
\label{x-temp}
\cos\theta=\partial_s x=\partial_s \kappa \, \partial_\kappa x=-\left(B \kappa_0\right)^{-1} n_\perp \partial_k x,
\end{equation}
and similarly for $y$. Knowing that $x=y=0$ at $k=1$, i.e. at $s=0$, and using Eqs.~(\ref{x-temp}) and (\ref{sol:n2}), the shape of the vesicle is thus given by the double quadrature
\begin{subequations}
\begin{align}
\label{x-y-eq}
\kappa_0\, x(k;K)&=\int_k^1\frac{\cos\theta(k';K)}{\mathcal{N}_\perp\left(k';\mathcal{P},K\right)}\,\mathrm{d} k', \\
\kappa_0\, y(k;K)&=\int_k^1\frac{\sin\theta(k';K)}{\mathcal{N}_\perp\left(k';\mathcal{P},K\right)}\,\mathrm{d} k'.
\end{align}
\end{subequations}
It turns out that, once Eq.~(\ref{eq:P(K0)}) is satisfied, $x$ automatically vanishes at the contact point, so that no new constraint results from that condition. Finally, the area of the vesicle is computed as
\begin{align}
A_1&=2\int_0^{y_\ell}x \,\mathrm{d} y=-2\int_{-1/K}^1x \left(\partial_k y\right) \mathrm{d} k \nonumber \\
&=\frac{2}{\kappa_0^2}\int_{-1/K}^1\frac{\kappa_0\, x(k;K) \sin\theta(k;K)}{\mathcal{N}_\perp\left(k;\mathcal{P},K\right)}\,\mathrm{d} k .
\end{align}
Having determined previously that $A_2/A=O(\Gamma^{-1})$, we have $A_1\simeq A$ and we obtain
\begin{subequations}
\label{eq:phi(K0)}
\begin{align}
\label{eq:phi(K0)-1}
\kappa_0^2A &\simeq \phi\left(K\right), \\
\label{eq:phi(K0)-2}
\phi\left(K\right)&=2\int_{-1/K}^1\frac{\kappa_0\, x(k;K) \sin\theta(k;K)}{\mathcal{N}_\perp\left(k;\mathcal{P},K\right)}\,\mathrm{d} k,
\end{align}
\end{subequations}
as long as $K\lesssim 3.9207$.
Like $\mathcal{P}(K)$, the function $\phi(K)$ is universal and independent of $\Gamma$, see Fig.~\ref{fig:P(K0)}(a). Using the result of Appendix~\ref{app:K392}, it is well fitted over all $K$ by
\begin{equation}
\phi(K)\approx \pi \left[\frac{1+3.373\, K^2+0.606\, K^4}{1+1.819\, K^2+0.276\, K^4}\right].
\label{fit:phi}
\end{equation}

\begin{figure*}
\centering
\includegraphics[width=\textwidth]{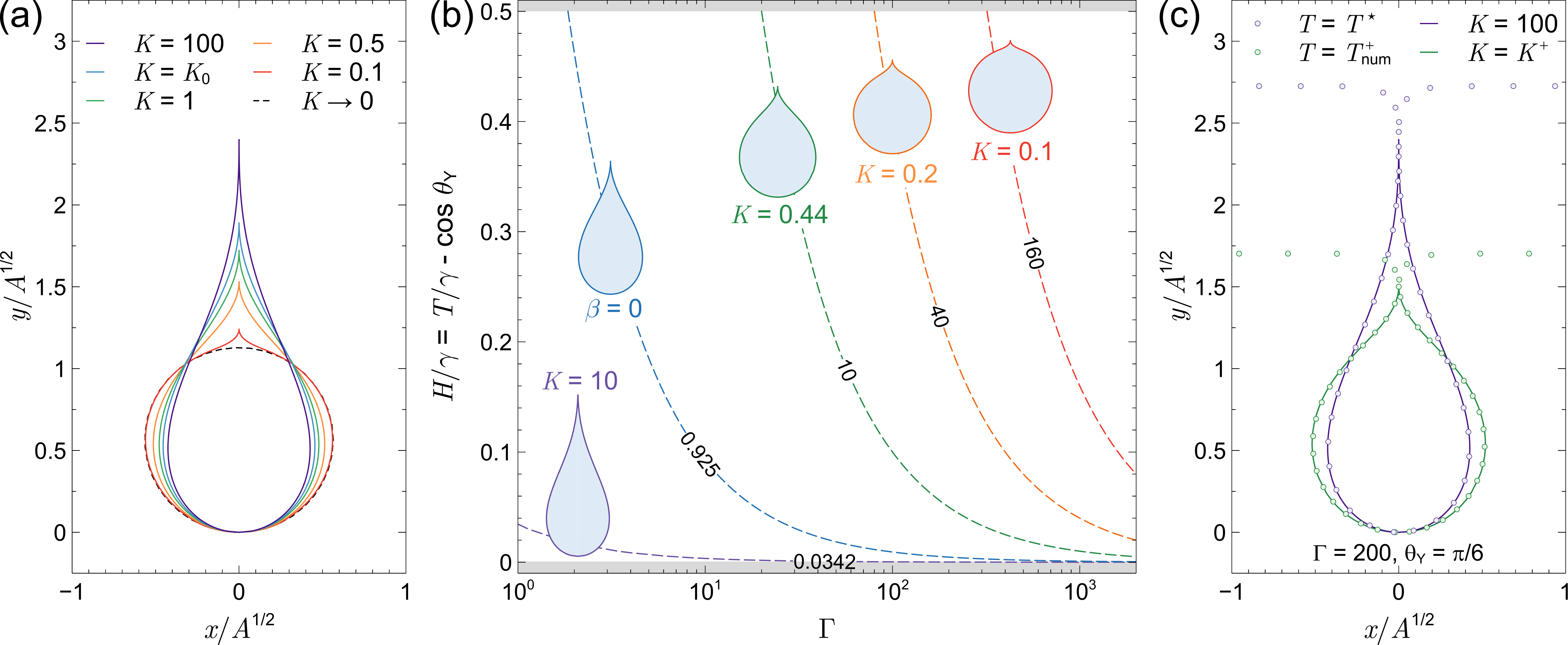}
\caption{(a) Family of vesicle shapes ($s \le \ell$) given by Eqs.~(\ref{eq:familly}) as a function of $K$. The applied tension increases as $K$ decreases and $K_0 \simeq 1.77842$ is the value for which $\beta = 0$. The vesicle shape approaches a circle of radius $(A/\pi)^{1/2}$ as $K\to 0$ and it barely changes when $K\gtrsim 10$. (b) Hyperbolas $H \, \Gamma/\gamma = c$ in the plane $(H/\gamma,\Gamma)$ along which the vesicle shape does not vary. The value of $c= \phi(K)/(2K^2)$ is indicated on the hyperbolas. (c) Comparison between vesicle shapes and those obtained numerically with $\Gamma=200$ and $\theta_Y = \pi/6$ for the extreme values of $T$. $T^{\star}/\gamma\simeq \cos \theta_Y$ is the numerical value of the tension below which the vesicle solution no longer exists and $T^{+}_{\text{num}}/\gamma \simeq 0.9114$ is the numerical value of the tension at which $F_c =0$. $K^{+}\simeq 0.447$ is obtained by solving $F_c =0$ using Eq.~(\ref{eq:Fc}) together with the expressions (\ref{Eq-T-beta}) of $\beta$ with $\Gamma=200$ and $\theta_Y = \pi/6$.}\label{fig:teardrops}
\end{figure*}

We are now able to simultaneously parametrise $\beta$ and $T$ with $K$. Indeed, using the last of Eqs.~(\ref{def:k,PP,K0}) and Eq.~(\ref{eq:phi(K0)-1}), we obtain $H$ as a function of $K$ and Eq.~(\ref{eq:defH2}) then gives $T$ as a function of $K$
\begin{align}
T(K)&=\gamma\cos\theta_Y+H(K), &H(K)=\gamma \frac{\phi\left(K\right)}{2\Gamma K^2}.
\label{phi/2K^2}
\end{align}
Finally, with $\beta \simeq \sin \beta$ in Eq.~(\ref{gammasinbeta/xD}), and having determined $\mathcal{P}(K)$, $H(K)$, and $T(K)$, we obtain
\begin{equation}
\label{Eq-T-beta}
\beta(K)\sim\mathcal{P}\left[\frac{H}{\gamma}\right]^{\frac{3}{2}}
\int_{\theta_Y}^{\pi/2}\frac{\cos\theta \, \mathrm{d} \theta}{\sqrt{H/\gamma +[1-T/\gamma]\cos\theta }}.
\end{equation}
Equations (\ref{phi/2K^2}) and (\ref{Eq-T-beta}) imply that $\beta=O(\Gamma^{-3/2})$ if $K=O(1)$ and that $\beta=O(\Gamma^{-1/2})$ if $H/\gamma=O(1)$, that is if $K=O(\Gamma^{-1/2})$. Hence, the assumption $\beta\ll 1$ made in deriving the solution above the contact point is verified in the large-$\Gamma$ limit. On the other hand, the radius of the liquid-vapour interface is obtained by using the second of Eqs.~(\ref{def:k,PP,K0}) and (\ref{phi/2K^2}) in Eq.~(\ref{laplace-pressure})
\begin{equation}
\frac{R_d}{\sqrt{A}}=\frac{2 \Gamma\, K^3}{\phi(K)^{3/2}\, \mathcal{P}(K)}.
\end{equation}
It is also interesting to compute the length $\ell(K)$ that makes half the perimeter of the vesicle. To this end, recall that  $\partial_s k = -n_\perp/\kappa_0 B$. Hence $\partial_k s=-\kappa_0 B/n_\perp$. From this, and assuming again that $K$ is less than 3.9207 so that $n_\perp$ does not vanish, we obtain
\begin{equation}
\ell(K) =\sqrt{\frac{A}{\phi(K)}} \int_{-1/K}^1\frac{\mathrm{d} k}{\mathcal{N}_\perp\left(k;\mathcal{P},K\right)} .
\end{equation}
The evolution of $\ell$ as a function of $K$ is shown in Fig.~\ref{fig:P(K0)}(b) together with some representative vesicle shapes.

Finally, the shape of the vesicle is obtained by combining Eqs.~(\ref{x-y-eq}) with Eq.~(\ref{eq:phi(K0)-1})
\begin{equation}
Z(k;K) =\sqrt{\frac{A}{\phi(K)}}\int_k^1\frac{e^{i\theta(k';K)}}{\mathcal{N}_\perp(k';\mathcal{P},K)}\,\mathrm{d} k',
\label{eq:familly}
\end{equation}
where $-1/K<k<1$ and $Z=x+iy$. For a given value of $K$, the above expression yields the shape of the vesicle, see Fig.~\ref{fig:teardrops}(a) and comparison to the numerics in Fig.~\ref{fig:teardrops}(c). Strikingly, this family of curves does not depend explicitly on $\Gamma$ and is therefore valid for arbitrary bending stiffness, provided that $\beta\ll1$. Nor does it depend on $\theta_Y$, which is understandable if there is no triple line within the vesicle. Additionally, the second of Eqs.~(\ref{phi/2K^2}) indicates that a given vesicle shape, identified by the single number $K$, is achieved over the locus of a constant product 
\begin{equation}
\frac{2H\Gamma}\gamma\equiv\frac{A}{\lambda_\text{BC}^2},
\end{equation}
where the product in question is $\phi(K)/K^2$, see Fig.~\ref{fig:teardrops}(b). 

Two particular vesicle shapes stand out in Fig.~\ref{fig:teardrops}(b). One is at $K=K_0$, where the capillary pressure vanishes. This implies that the liquid-vapour interface is flat, i.e. that $\beta=0$. This corresponds to the unique curve $Z(k,K_0)$, with $-1/K_0<k<1$, in agreement with the numerical curves of Fig.~\ref{fig-shapes}(e). Interestingly, when the capillary pressure vanishes, the shape of the vesicle is the same as in the absence of the fluid. Hence, we expect the curve $Z(k,K_0)$ to describe the `self-encapsulation' state of the `dripping' elastic rod described in~\cite{Bosi2015}. The second corresponds to the limit $K\to\infty$, where $H \to 0$ so that $T\to \gamma \cos\theta_Y$ and $\kappa_\ell\to 0$, see Eqs.~(\ref{eq:defH3}) and (\ref{phi/2K^2}). For lower tensions, i.e. $T<T^{\star} \equiv \gamma \cos\theta_Y$, the vesicle does not exist. The fact that $\kappa_\ell$ tends to zero suggests that the disappearance of the vesicle state occurs through a lengthening and thinning of the region of contact. 

\subsection{Contact force}

The existence of the vesicle state requires that $F_c>0$. The tension $T^+$ for which $F_c$ vanishes is also the bifurcation point with the partial wetting solution. From Eqs.~(\ref{eq:nperp1bis}), we have $F_c(K)=\lim_{\epsilon\to0}\left[n_\perp(\ell+\epsilon)-n_\perp(\ell-\epsilon)\right]$. The expression of $n_\perp$ at $s=\ell^+$ is obtained from Eq.~(\ref{eq:nperp2bis}) with $\theta_D = \theta_Y-\beta$. The expression of $n_\perp$ at $s=\ell^-$ is obtained by using Eqs.~(\ref{sol:n2-def:NN}) with $k=-1/K$, Eq.~(\ref{eq:phi(K0)-1}) and the first of Eqs.~(\ref{phi/2K^2}). The expression of the contact force as a function of $K$ reads then
\begin{subequations}
\begin{align}
\label{eq:Fc}
\frac{F_c(K)}{\gamma} &= \frac{\cos\theta_Y}{\cos\left(\theta_Y-\beta\right)}-\cos\theta_Y-\frac{\mathcal{F}(K)}{\Gamma}, \\
\label{FF-def}
\mathcal{F} (K)&=\frac{\phi(K)}{2K^2}\left[1+\sqrt{4\left(1+K\right)\mathcal{P}+\left(K^2-1\right)^2}\right].
\end{align}
\end{subequations}
Numerically solving $F_c(K)=0$ yields the root $K^+$ and therefore, from Eqs.~(\ref{phi/2K^2}), $T^+$ as a function of $\theta_Y$ and $\Gamma$, in very good agreement with the numerical simulations (see the curve $T^+_{\text{a}}$ in Fig.~\ref{fig-phase-diagram}). In the large-$\Gamma$ limit, we have $\beta\ll1$ and $K^+=O(\Gamma^{-1/2})$. It is easy to find, with the aid of Eqs.~(\ref{fit:PP}) and (\ref{fit:phi}), that 
\begin{equation}
K^+ \simeq \left(\frac{\pi/\Gamma}{ 1-\cos\theta_Y}\right)^{1/2},
\end{equation}
corresponding to $T^{+} \simeq \cos^2(\theta_Y/2)$.
A more detailed calculation yields 
\begin{equation}
T^{+} \simeq \cos^2(\theta_Y/2)-C^{+}\left(\theta_Y\right)\Gamma^{-1/2},
\end{equation}
where $C^{+}\left(\theta_Y\right)$ is a complicated function that is approximated in the range $0<\theta_Y<\pi/2$ by
\begin{equation}
C^+\approx \frac{\sqrt{\pi/8}\,\theta_Y}{1-0.296\, \theta_Y+0.235\, \theta_Y^2}.
\end{equation}

\subsection{Limiting shape at $T=T^\star$}
\label{shape-t=ts}

We close this section by noting that, as $T\to T^\star$, the boundary conditions of Eqs.~(\ref{eq:nperp1}) in the range $|s|<\ell$ are $\kappa=n_\parallel=0$ at $s=\ell$, so that the problem is mathematically equivalent to the one studied by Mora \textit{et al.} in Ref.~\cite{Mora2012} using the method of Ref.~\cite{Djondjorov2011} (with corrected boundary conditions). The study in Ref.~\cite{Mora2012} addressed the shape of a fishing line deformed by the surface tension of a soap film. At the particular point $T=T^{\star}$, corresponding to $K\to\infty$ in our theory, the vesicle shape is thus given, up to a scale factor, by the solution explicitly given in Ref.~\cite{Mora2012}. A similar shape is also found in portions of solutions reported in Refs.~\cite{Djondjorov2011,Py07}. In Appendix~\ref{app:racket}, we provide an alternative formulation of the solution reported in Ref.~\cite{Mora2012} based on the present theory.

\section{Conclusions and perspectives}\label{sec:conclusion}

In this paper, we have presented a comprehensive picture of the bending of a thin elastic sheet under the opposite actions of capillary forces and an external tension $T$. In order to elucidate the essential mechanisms, we have focused on a sheet that is much larger than the drop size, so that it can be modelled as being infinitely long. When $T=0$, the system has been used as a 2D model for ``capillary origamis'' and configurations of complete wetting had already be reported in this context~\cite{Py07}. However, applying an external tension significantly modifies the folding dynamics. Now, the conformation of the system depends on $T$ and upon varying this parameter, we found the possibility of wrapping most of the liquid inside a vesicle, which corresponds to the ``budding transition'' described by Kusumaatmaja and Lipowsky~\cite{Kusumaatmaja2011}. In this regard, one of the most dramatic results of our study is that, at the vicinity of the transition from the wrapped vesicle state to the complete wetting, the vesicle shape is universal -- being independent on the explicit value of the bending rigidity $B$, but nevertheless distinct from a the circular shape that is obtained for $B=0$. This is surprising for one may expect a small $B$ to manifest itself only in boundary layers whose size is comparable to the bendocapillary length $\ell_\text{BC}=\sqrt{B/\gamma}$. One intuitive explanation for this, which is motivated by the asymptotic analysis of Sec.~\ref{sec:partialwrap}, is that the effective bendocapillary length is not $\ell_\text{BC}$ but rather the tension-dependent length scale $\lambda_{\text{BC}}=\sqrt{B/2(T-\gamma\cos\theta_Y)}$ . As the denominator of this expression tends to zero, the balance of bending and capillary forces is pronounced in the whole vesicle. 
 
One of the questions that motivated the introduction of a finite $B$ in the model was whether this would induce a snapping transition between the partial wetting and vesicle states, i.e. would the former emerge subcritically from the latter. The answer is yes, but not in the asymptotic limit $\Gamma\gg1$: numerically, there is a finite value $\Gamma^{+}(\theta_Y)$ below which the transition becomes subcritical. For small $\theta_Y$, we find that $\Gamma^{+}(\theta_Y)$ scales as $\theta_Y^{-2}$. This observation suggests that the limit of a small $\theta_Y$ is singular. While we have not studied the double limit $\Gamma\gg1$ and $\theta_Y\ll1$, analyzing this asymptotic regime may enable one to analytically capture the subcritical vesicle-partial wetting transition. 

Numerical simulations with lower values of $\Gamma$ unexpectedly revealed that the partial wetting state can exist at applied tensions $T$ significantly smaller than the threshold for complete wetting $T^\star=\gamma\cos\theta_Y$, see Fig.~\ref{fig-bifurcation-diag-numerics}\,(a). In that scenario, the partial wetting state disappears upon decreasing $T$ without exhibiting the vesicle state. It gives way, at a limit point $T^-$ close to $T=0$, to the complete wetting state. If, subsequently, $T$ is increased, one expects the vesicle state to emerge from the complete wetting state at $T^\star$. Further, at $T^+$, the vesicle state opens and the system jumps discontinuously to the partial wetting state.

For an infinitely long sheet, the complete wetting state can be realized only after infinitely long time. That is, if $T<T^\star$, the vesicle state disappears and the tension is not sufficient to counteract capillary forces. Hence, an infinitely long stretch of sheet is entrained by capillarity, in a never-ending process. Consistently with this, $T^\star$ is precisely the value at which the curvature at self-contact vanishes in the vesicle state. Geometrically, this allows the self-contact point to become a segment of line of arbitrary length. Interestingly, the corresponding limiting shape coincides with that of a fishing line or hair that collapses onto itself when dipped in a soapy solution~\cite{Mora2012}. This is a particular case in our theory and we thus provide an alternative formula for what these authors call a ``tennis racket'' loop, see Appendix~\ref{app:racket}.

We have not studied Young's angles in the range $\pi/2<\theta_Y<\pi$. The threshold for complete wetting, $T^\star=\gamma\cos\theta_Y$ suggests that in that case a complete wetting state may be realized only if the sheet is under compression. This amounts to completely modify the mechanics of the problem: for one thing, an infinitely long sheet would buckle at arbitrarily small compressive stresses. Thus, considering this range of Young's angle would require us to abandon our simplifying hypothesis and include the sheet length, $L$, as a key parameter. These two aspects, $\pi/2<\theta_Y<\pi$ and finite $L$, open interesting research perspectives on this basic physical setting.

To close this conclusion, we must mention the work by Kusumaatmaja and Lipowski~\cite{Kusumaatmaja2011}, who numerically studied a (3D) axisymmetric bud forming in a membrane under tension and in contact with two distinct fluids. This budding solution is analogous to the vesicle solution described in the present paper (the authors studied it as a function of the nondimensionalised drop volume, i.e. in the present notation, as a function of $\Gamma^{3/2}$). However, the presence of hoop stress prevented them from obtaining analytical results for $B>0$ and, in this sense, the present work provides some analytical support to Ref.~\cite{Kusumaatmaja2011}.

\section*{Acknowledgement}

The research leading to these results has received funding from NSF-CAREER Grant No. DMR 11-51780 (BD), visitor grant from the Fonds de la Recherche Scientifique - FNRS (BD) and W. M. Keck Foundation (BD, FB). GK is a Research Associate of the Fonds de la Recherche Scientifique - FNRS.

We are grateful to R. Govindarajan, J. Hanna, N. Menon, S. Neukirch, S. Walker and D. Vella for useful discussions. BD benefited from stimulating discussions with participants of the program ``Geometry, elasticity, fluctuations, and order in 2D soft matter'', held in winter 2016 at the Kavli Institute for Theoretical Physics, UCSB.

\appendix

\section{Minimization of the Lagrangian~(\ref{lagrangian-no-B})}
\label{app-mini}

We give here details about the minimization of the Lagrangian~(\ref{lagrangian-no-B}) leading to the Eqs.~(\ref{eqs-equi}). Requiring that the derivatives of $\mathcal{L}$ with respect to $\beta$, $\vartheta$, $R_d$ and $L_b$ vanishes, leads to respectively
\begin{subequations}
\begin{align}
\label{eq-L-beta}
2\gamma\, R_d - 2R_d\, T\, \cos \beta - \mu \frac{\partial \mathcal{A}}{\partial \beta} + 2\eta\, R_d \frac{\vartheta \cos \beta}{\sin \vartheta} &= 0, \\
\label{eq-L-vartheta}
-\mu \frac{\partial \mathcal{A}}{\partial \vartheta} +2\eta \,R_d \frac{\sin \beta}{\sin \vartheta}\left(1-\frac{\vartheta\, \cos \vartheta}{\sin \vartheta}\right) &= 0, \\
\label{eq-L-Rd}
2\gamma\, \beta - 2T\, \sin \beta-\mu \frac{\partial \mathcal{A}}{\partial R_d}+2\eta \frac{\vartheta\, \sin \beta}{\sin \vartheta} &=0, \\
\label{eq-L-Lb}
-\Delta \gamma + T - \eta &= 0.
\end{align}
\end{subequations}
Equation (\ref{eq-L-Lb}) gives immediately
\begin{equation}
\eta \equiv n_{\parallel}= T-\Delta \gamma,
\end{equation}
which is just Eq.~(\ref{eqs-eta2}). Using the expression~(\ref{nobending:eq:A}) of $\mathcal{A}(\beta,\vartheta,R_d)$, we find that
\begin{equation}
\frac{\partial \mathcal{A}}{\partial \vartheta} = \frac{2R_d^2\, \sin^2\beta}{\sin^2\vartheta}\left(1-\frac{\vartheta\, \cos \vartheta}{\sin \vartheta}\right).
\end{equation}
Therefore, Eqs.~(\ref{eq-L-vartheta}) gives a relation between the two Lagrange multipliers $\mu$ and $\eta$
\begin{equation}
\label{eqs-eta}
\eta = \mu\, R_d \frac{\sin \beta}{\sin \vartheta}.
\end{equation}
Multiplying Eq.~(\ref{eq-L-beta}) by $\sin \beta$ and Eq.~(\ref{eq-L-Rd}) by $R_d \cos \beta$ and subtracting the resulting equations, we have
\begin{align}
\label{temp1}
2 \gamma\, R_d(\sin \beta - \beta\, \cos \beta) \nonumber &-\mu \left[\frac{\partial \mathcal{A}}{\partial \beta}\, \sin \beta \right. \nonumber \\ &- \left. R_d \frac{\partial \mathcal{A}}{\partial R_d}\, \cos \beta \right] = 0.
\end{align}
Using Eq.~(\ref{nobending:eq:A}) we find that
\begin{equation}
\frac{\partial \mathcal{A}}{\partial \beta}\, \sin \beta - R_d \frac{\partial \mathcal{A}}{\partial R_d}\, \cos \beta = 2R_d^2\, (\sin \beta - \beta\, \cos \beta).
\label{temp2}
\end{equation}
Substituting Eq.~(\ref{temp2}) into Eq.~(\ref{temp1}) leads to the expression of the Lagrange multiplier $\mu$
\begin{align}
\label{eq-mu-eta}
\mu \equiv p &= \frac{\gamma}{R_d}, &\eta &\equiv n_{\parallel} = \gamma \frac{\sin \beta}{\sin \vartheta}=\gamma \frac{R_b}{R_d},
\end{align}
where we used Eqs.~(\ref{eqs-eta}) and (\ref{geometry-zero-B}) in the second equation. We thus recover the first of Eqs.~(\ref{eqs-eta1}). Eqs.~(\ref{eq-mu-eta}) reveal the Lagrange multipliers $\eta$ and $\mu$ as the tension $n_\parallel$ in the wet part of the sheet and the pressure $p$ in liquid volume, respectively. Substituting Eqs.~(\ref{eq-mu-eta}) into Eqs.~(\ref{eq-L-Rd}) (or Eq.~(\ref{eq-L-beta})), we have
\begin{equation}
2\gamma\, \beta - 2T\, \sin \beta-2\gamma \frac{\mathcal{A}}{R_d^2}+2\gamma \frac{\vartheta\, \sin^2 \beta}{\sin^2 \vartheta}=0,
\end{equation}
where we used $\partial \mathcal{A}/\partial R_d = 2\mathcal{A}/R_d$. Finally, using Eq.~(\ref{nobending:eq:A}), we obtain
\begin{equation}
\label{force-balance}
T=\gamma\, \cos \beta + \gamma \frac{\sin \beta \cos \vartheta}{\sin \vartheta} = \gamma\, \cos \beta + n_{\parallel} \, \cos \vartheta,
\end{equation}
which coincide with the second of Eqs.~(\ref{eqs-eta1}).

\section{Additional numerical results}
\label{app-num}

\begin{figure*}[!t]
\centering
\includegraphics[width=\textwidth]{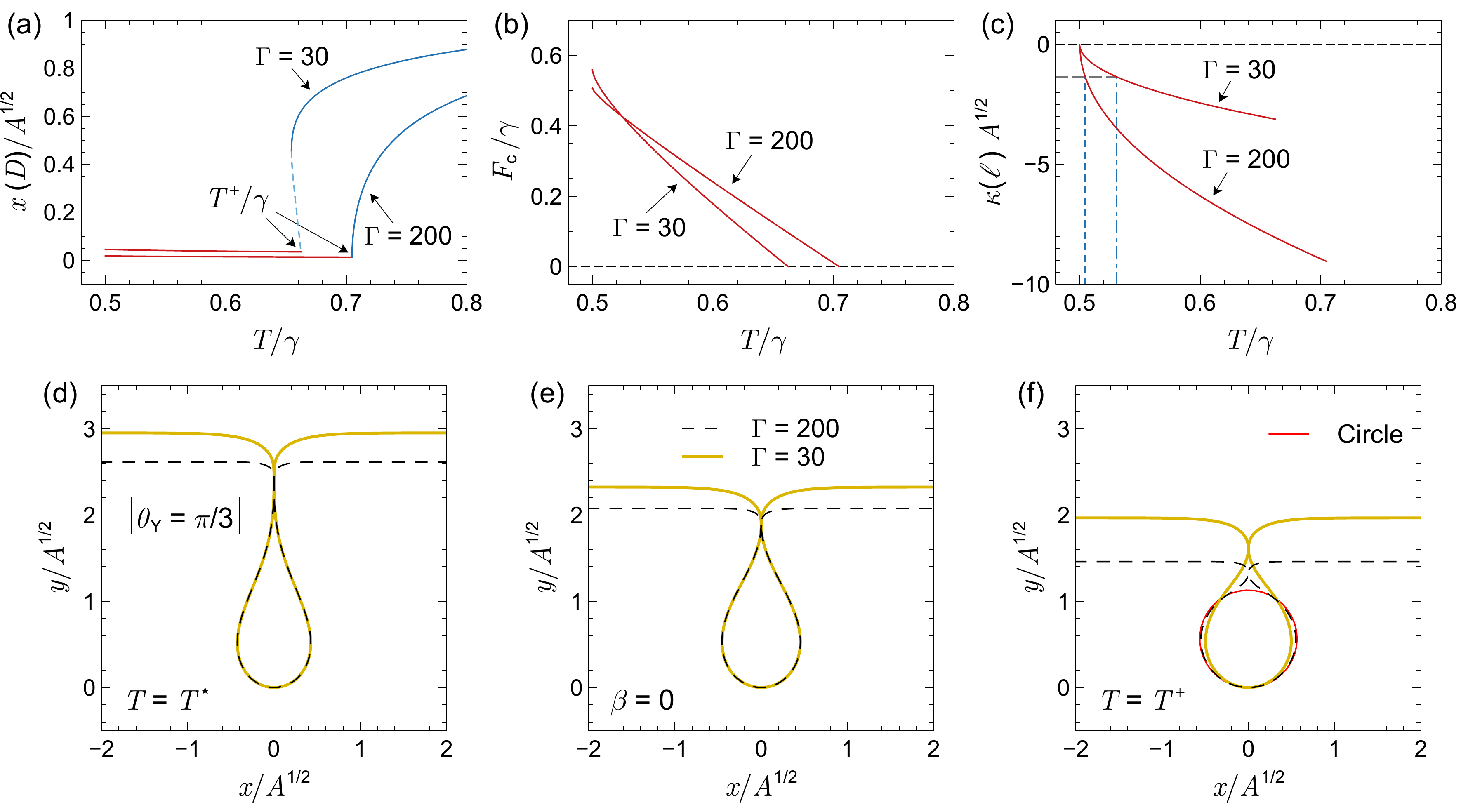}
\caption{(a)-(c) Evolution of various quantities characterising the system shape as a function of $T/\gamma$ for $\theta_Y = \pi/3$ and $\Gamma = 30$ and $200$. As in Fig.~\ref{fig-bifurcation-diag-numerics}, the blue solid and dashed curves correspond respectively to the stable and unstable branches of the partial wetting state whereas the red curves refer to the vesicle state. Panel (c) shows that $\kappa(\ell)$ is independent on $\Gamma$ when $T=T^{\star}=\cos \theta_Y$ and $\beta=0$. (d)-(f) Influence of $\Gamma$ on the vesicle shapes at three different values of $T$ for $\theta_Y=\pi/3$. The shape of the vesicle is independent on $\Gamma$ when $T\to T^{\star}$ and $\beta=0$ (in agreement with panel (c)) whereas it does depend on this parameter at $T=T^{+}$ where the shape approaches a circular shape of radius $R=(A/\pi)^{1/2}$ as $\Gamma$ increases, see panel (f).
}
\label{fig-shapes}
\end{figure*}

Figure~\ref{fig-shapes}(a)-(c) shows the evolution of various quantities characterizing the system shape as a function of $T$ for two values of $\Gamma$ and $\theta_Y=\pi/3$. When $\Gamma$ increases, the position of the contact point along the $x$-axis, $x_D$, tends to zero at the transition between partial wetting and the vesicle state, i.e. at $T=T^{+}$, and stays small in the vesicle state, see Fig.~\ref{fig-shapes}(a). This is consistent with the observation that the length of the sheet forming the vesicle, $\ell$, is close to $D$. Therefore, essentially all the liquid is encapsulated in the vesicle as $\Gamma$ increases. The contact force, $F_c$, vanishes at $T = T^{+}$ and increases almost linearly when the applied tension decreases and reaches a value $\gamma-\gamma\cos \theta_Y$ at $T = T^{\star}$ when $\Gamma \to \infty$, as shown by the asymptotic theory presented in Sec.~\ref{sec:partialwrap} (see Fig.~\ref{fig-shapes}(b)).

Figure~\ref{fig-shapes}(c) shows that $\kappa_\ell$ is independent of $\Gamma$ at $T=T^{\star}$. It also shows that the tension at which $\beta =0$ depends on $\Gamma$ in agreement with the results reported in Fig.~\ref{fig-bifurcation-diag-numerics}. However, the values of $\kappa_\ell$ at $\beta =0$ is again independent on $\Gamma$. This suggests that the vesicle shape does not depend on $\Gamma$ for some particular values of the tension as confirmed by Fig.~\ref{fig-shapes}(d)-(e). However, as shown in Fig.~\ref{fig-shapes}(f), the vesicle shape at $T = T^{+}$ does depend on $\Gamma$ and approaches a circular shape of radius $(A/\pi)^{1/2}$ as $\Gamma \to \infty$ (see also Fig.~\ref{fig:teardrops}). This striking observation is fully explained by the asymptotic theory, see Sec.~\ref{sec:partialwrap}, which shows that the vesicle has a given shape when $(T-\gamma\cos \theta_Y)\Gamma$ is constant. This is obviously the case when $T=T^{\star}\equiv \cos \theta_Y$ and the asymptotic theory shows that this is also the case when $\beta=0$. However, the product $(T-\gamma\cos \theta_Y)\Gamma$ and, hence, the vesicle shape, does change with $\Gamma$ at $T = T^{+}(\theta_Y,\Gamma)$.

\section{Algorithm to find $\Gamma_c^{+}$}\label{app:gammac}

By definition, $\Gamma^+(\theta_Y)$ is the value of $\Gamma$ at which the self-contact state with a vanishing contact force ($F_c=0$) switches from the unstable to the stable branch of the partial wetting state at a given $\theta_Y$. Self-contact solutions of Eqs.~(\ref{ODE-general-p-wet}) can be computed by adding two additional shooting parameters, $T$ and $\ell$, which are fixed thanks to two additional boundary conditions, $\theta(\ell)=0$ and $x(\ell)=0$. The particular value of the tension at which such a state is found is, by definition, $T^{+}$. This procedure allows the self-contact state with a vanishing contact force to be computed for given values of $\Gamma$ and $\theta_Y$.

To determine $\Gamma^+$ numerically, $\Gamma$ is increased by small steps for a given value of $\theta_Y$. When $\Gamma < \Gamma^+$, the self-contact state belongs to the unstable branch of the partial wetting state and the transition is subcritical and, when $\Gamma > \Gamma^+$, it belongs to the stable branch and the transition is supercritical. 

To determine at which branch the self-contact state belongs to, the self-contact solution is perturbed at each value of $\Gamma$ by slightly increasing $T$, \textit{i.e.} $T=T^{+} + \Delta T$ with $0<\Delta T/T^{+}\ll 1$. If the self-contact solution belongs to the unstable branch, then the perturbed solution will feature some self-crossing, i.e. $\min_s x(s)$ of the perturbed solution near $s=\ell$ is negative. If $\min_s x(s)>0$, then the self-contact solution belongs to the stable branch. At a given $\theta_Y$, $\Gamma^+$ corresponds thus to the value of $\Gamma$ at which $\min_s x(s)$ changes its sign. 

\section{Vesicle with $K\gtrsim 3.9207$}\label{app:K392}

In Sec.~\ref{sec:Kless392}, we assumed that $\kappa$ decreases monotonously from $\kappa_0>0$ to $\kappa_\ell<0$. This assumption ceases to hold for the range $3.9207\lesssim K<\infty$. In that range of values, $n_\perp$ necessarily vanishes somewhere along the curve and the formulas of Sec.~\ref{sec:Kless392} must be revised. The change of sign of $\partial_s\kappa$ happens when $\mathcal{N}_\perp$ vanishes, that is at $k=k_\text{min}$, solution of
\begin{equation}
\frac{\mathcal{P}}{K^3}+\frac{1+k_\text{min}}{4}\left(1+k_\text{min}^2-\frac{2}{K^2}\right)=0.
\label{eq:kmin}
\end{equation}
Since the above equation is of third order, a closed form expression can be written for $k_\text{min}$ in terms of $\mathcal{P}$ and $K$:
\begin{equation}
k_\text{min}= -\frac{1}{3}+\frac{ W^{1/3}}{3}-\frac{2}{ W^{1/3}}\left(\frac1{3}-\frac{1}{K^2}\right),
\end{equation}
where
\begin{align}
\frac{WK^3}{2} &=9K-5K^3-27\mathcal{P} +\sqrt{27}\left[-2+5K^2 \right.\\
&- \left. 4K^4+K^6-18K\mathcal{P}+10K^3\mathcal{P}+27\mathcal{P}^2 \right]^{1/2}.  \nonumber 
\end{align}
Starting from the lowermost point of the vesicle, \textit{i.e.} $x=y=s=0$, $k$ first decreases from 1 to $k_\text{min}<0$, before increasing again from $k_\text{min}$ to $-1/K$. Starting from $k=1$ and as long as $n_\perp>0$, the function $\theta(k;K)$ has the expression [see Eq.~(\ref{def-theta})]
\begin{align}
\theta_1(k)&=\int_{k}^1 \frac{k'\,\mathrm{d} k'}{\mathcal{N}_\perp\left(k';\mathcal{P},K\right)},
&k_\text{min}<k<1
\label{app:theta1}
\end{align}
Once $k_\text{min}$ is reached, $n_\perp$ changes sign and, subsequently,
\begin{align}
\theta(k;K)=\theta_2(k)=\theta_1(k_\text{min})+\int_{k_\text{min}}^{k} \frac{k' \,\mathrm{d} k'}{\mathcal{N}_\perp\left(k';\mathcal{P},K\right)},
\end{align}
where $k_\text{min}<k<-1/K$ and  $\mathcal{N}_\perp$ is still the positive function defined in Eq.~(\ref{def:NN}). Note that
$
\theta_2(k)=2\theta_1(k_\text{min};K)-\theta_1(k),
$
so that only $\theta_1(k)$ needs to be evaluated in practice. The equation that yields $\mathcal{P}(K)$ is now
\begin{equation}
\int_{k_\text{min}}^1 \frac{k \,\mathrm{d} k}{\mathcal{N}_\perp\left(k;\mathcal{P},K\right)}+\int_{k_\text{min}}^{-1/K} \frac{k\,\mathrm{d} k}{\mathcal{N}_\perp\left(k;\mathcal{P},K\right)} =\frac{\pi}2.
\end{equation}

\begin{figure}[!t]
\centering
\includegraphics[width=\columnwidth]{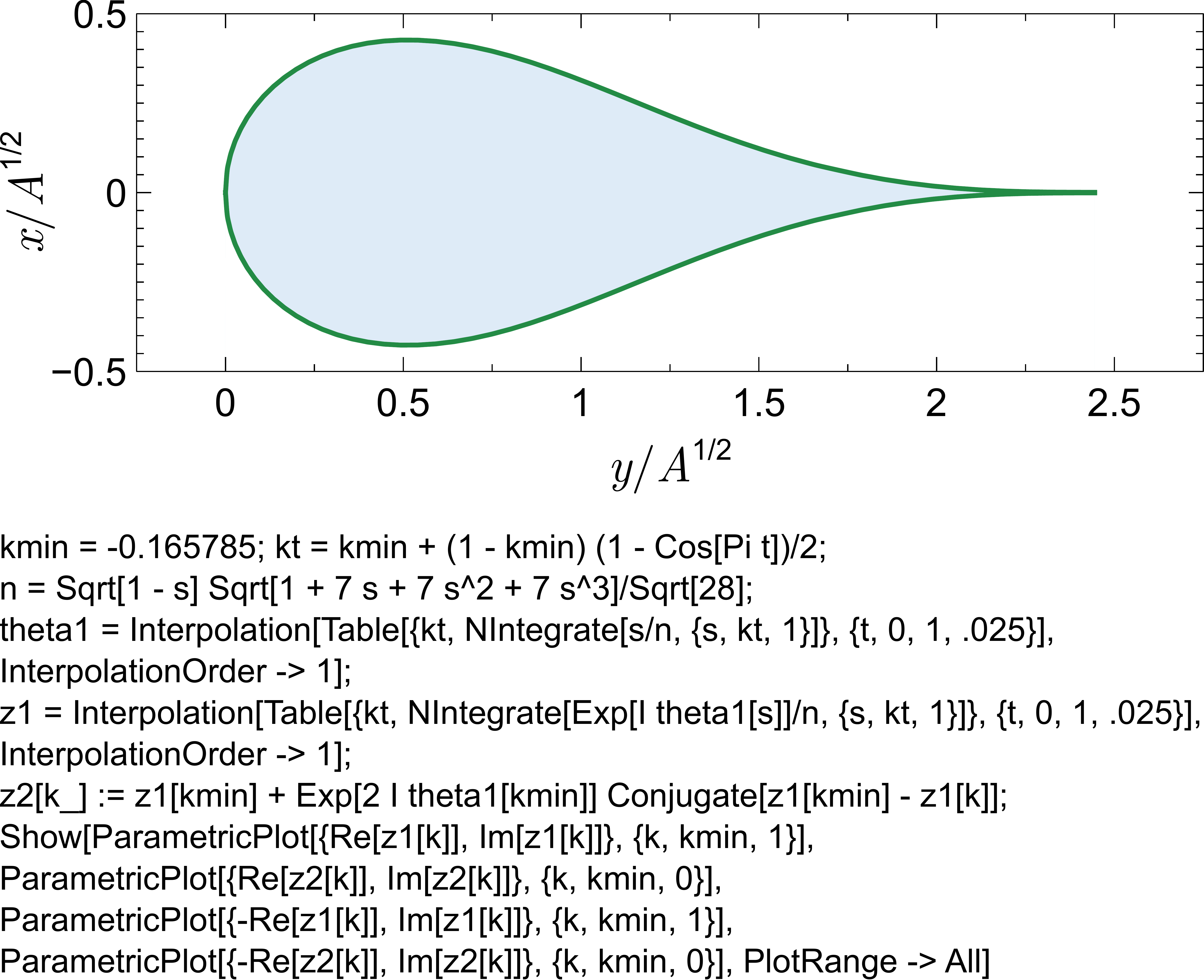}
\caption{Limiting vesicle shape at the verge of complete wetting, and the Mathematica code used to draw it. This is the same shape as the ``tennis racket'' solution in Ref.~\cite{Mora2012}.}\label{fig:racket}
\end{figure}

Once $\mathcal{P}(K)$ is determined, we may compute the complex coordinates
\begin{align}
\eta_1(k)&=\int_{k}^1\frac{e^{i\theta_1(k')}\, \mathrm{d} k'}{\mathcal{N}_\perp\left(k';\mathcal{P},K\right)}, &k_\text{min}&<k<1,
\label{app:eta1}
\end{align}
of which $\kappa_0 x_{1}(k)$ and $\kappa_0 y_{1}(k)$ are the real and imaginary parts, respectively [see Eq.~(\ref{x-y-eq})]. Similarly,
\begin{align}
\eta_2(k)=\eta_1(k_\text{min})+\int_{k_\text{min}}^k\frac{e^{i\theta_2(k')}\, \mathrm{d} k'}{\mathcal{N}_\perp\left(k';\mathcal{P},K\right)},
\end{align}
where $k_\text{min}<k<-1/K$. Only the function $\eta_1(k)$ needs to be evaluated, for we have
\begin{equation}
\eta_2(k)= \eta_1(k_\text{min})+e^{2i\theta_1(k_\text{min})}\left[\eta_1^*(k_\text{min}) -\eta_1^*(k)\right],
\label{app:eta2}
\end{equation}
where $\eta_1^*$ is the complex conjugate of $\eta_1$. Combining the expressions just obtained, one may derive
\begin{align}
\phi(K) &=A\kappa_0^2=-2\int_{k_\text{min}}^1\Re[\eta_1(k)] \Im[\eta_1'(k)] \mathrm{d} k \nonumber \\
&+2\int_{k_\text{min}}^{-1/K} \Re[\eta_2(k)] \Im[e^{2i\theta_1(k_\text{min})}\eta_1'^*(k)] \mathrm{d} k,
\end{align}
where $\Re[\cdot]$ and $\Im[\cdot]$ denote real part and imaginary part, respectively, and  $\eta_1'(k)$ is the derivative of $\eta_1$ with respect to $k$.
To close this section, let us compute the length of the curve that makes up the vesicle. One has $\partial_s k = -n_\perp/\kappa_0 B$. Hence $\partial_k s=-\kappa_0 B/n_\perp$. From this, and bearing in mind the change of sign of $n_\perp$, one obtains
\begin{align}
\ell(K) &=\sqrt{\frac{A}{\phi(K)}}\left(\int_{k_\text{min}}^1\frac{\mathrm{d} k}{\mathcal{N}_\perp\left(k;\mathcal{P},K\right)} \right. \nonumber \\
&+ \left. \int_{k_\text{min}}^{-1/K}\frac{\mathrm{d} k}{\mathcal{N}_\perp\left(k;\mathcal{P},K\right)}\right).
\end{align}

\section{The ``tennis racket'' shape}\label{app:racket}

We conclude by giving the solution as $K\to\infty$ ($T \to T^{\star}$), which is an alternative formulation of the solution of Ref.~\cite{Mora2012}. In that limit, $\mathcal{P}\sim-3K^3/14$ and
\begin{subequations}
\begin{align}
\mathcal{N}_\perp &\to\sqrt{1-k}\sqrt{1+7k+7k^2+7k^3}/\sqrt{28}, \\
W &\to\left(11+3\sqrt{57}\right)/7, \quad k_\text{min}\to-0.165785\ldots
\end{align}
\end{subequations}
The profile shown in Fig.~\ref{fig:racket} is obtained by evaluating Eq.~(\ref{app:theta1}) together with the real and imaginary parts of the functions (\ref{app:eta1}) and (\ref{app:eta2}) and using $\phi_m=\phi(K\to\infty)\simeq 6.89495$. Note that, contrary to Ref.~\cite{Mora2012}, no root-finding is necessary to obtain the solution.


\end{document}